\documentclass[traditabstract,longauth]{aa} 
\usepackage[american]{babel}
\usepackage{graphicx}
\usepackage{txfonts}
\usepackage{natbib}
\bibpunct{(}{)}{;}{a}{}{,}
\def\lsim{\,\lower2truept\hbox{${< \atop\hbox{\raise4truept\hbox{$\sim$}}}$}\,}
\def\gsim{\,\lower2truept\hbox{${> \atop\hbox{\raise4truept\hbox{$\sim$}}}$}\,}
\begin{document}
\title{Planck pre-launch status: the Planck-LFI programme}
\author{N.~Mandolesi$^1$,
M.~Bersanelli$^2$,
R.C.~Butler$^1$,
E.~Artal$^7$,
C.~Baccigalupi$^{8,35,6}$,
A.~Balbi$^5$,
A.J.~Banday$^{9,39}$,
R.B.~Barreiro$^{17}$,
M.~Bartelmann$^9$,
K.~Bennett$^{26}$
% ...
}
\institute{
INAF -- IASF Bologna, Istituto Nazionale di Astrofisica, Istituto di Astrofisica Spaziale e Fisica Cosmica di Bologna, Via Gobetti 101, I-40129 Bologna, Italy
  \and
 Dipartimento di Fisica, Universit\`a degli Studi di Milano, Via Celoria, 16, 20133 Milano, Italy
   \and
 INAF -- OAPd, Istituto Nazionale di Astrofisica, Osservatorio Astronomico di Padova, Vicolo dellÔOsservatorio 5, I-35122 Padova, Italy
   \and
 Dipartimento di Fisica ÒG. GalileiÓ, Universit\`a degii Studi di Padova, via Marzolo 8, I-35131 Padova, Italy
   \and
    Dipartimento di Fisica, Universit\`a degli Studi di Roma ``Tor Vergata'', via della Ricerca Scientifica 1, 00133 Roma, Italy
   \and 
   INAF -- OATs, Istituto Nazionale di Astrofisica, Osservatorio Astronomico di Trieste, Via G.B.~Tiepolo 11, I-34131, Trieste, Italy
  \and
  Dep. Ing. de Comunicaciones (DICOM), Universidad de Cantabria Av. De Los Castros S/N, 39005 Santander, Spain
  \and  
   SISSA/ISAS, Scuola Internazionale di Studi Superiori Avanzati/International Schools for Advanced Studies, Astrophysics Sector, via Beirut 2-4, Sezione di Trieste, I-34014, Trieste, Italy
  \and
  MPA -- Max-Planck-Institut f\"ur Astrophysik, Karl-Schwarzschild-Str. 1, 85741 Garching, Germany
  \and
Jet Propulsion Laboratory, California Institute of Technology
4800 Oak Grove Drive, Pasadena, CA 91109, USA
  \and
  ISDC Data Centre for Astrophysics, University of Geneva, ch. d'Ecogia 16, CH-1290 Versoix, Switzerland
   \and
  {\it Herschel}/{\it Planck\/} Project, Scientific Projects Dpt of ESA, Keplerlaan 1, 2200 AG, Noordwijk, The Netherlands
  \and
  IFP-CNR, Istituto di Fisica del Plasma, Consiglio Nazionale delle Ricerche, Via Roberto Cozzi, 53, 20125 Milano, Italy
  \and
  Jodrell Bank Centre for Astrophysics, University of Manchester, M13 9PL, UK
  \and
  ASI, Agenzia Spaziale Italiana, Viale Liegi, 26, 00198 Roma, Italy
 \and
 Dipartimento di Fisica, Universit\`a di Trieste, via A. Valerio n.2 - 34127 Trieste, Italy
 \and
 Instituto de Fisica de Cantabria, CSIC- Universidad de Cantabria, Avenida de los Castros s/n, 39005 Santander, Spain
 \and
 Instituto de Astrof\'isica de Canarias, C/ V\'ia L\'actea s/n, 38200, 
La Laguna, Tenerife, Spain
 \and
 University of Helsinki, Department of Physics,
P.O. Box 64, FIN-00014 Helsinki, Finland
 \and
Mets\"{a}hovi Radio Observatory, TKK, Helsinki University of Technology,
Mets\"{a}hovintie 114, FIN-02540 Kylm\"{a}l\"{a}, Finland
 \and
 Physics Department, University of California, Santa Barbara, CA 93106, USA
 \and
 Institute of Theoretical Astrophysics, University of Oslo, P.O. Box 1029 Blindern, N-0315 Oslo, Norway 
 and 
 Centre of Mathematics for Applications, University of Oslo, P.O. Box 1053 Blindern, N-0316 Oslo, Norway
 \and
 ESA/ESAC/RSSD, European Space Agency, European Space Astronomy Centre, Research and Scientific Support Department,  P.O. Box - Apdo. de correos 78, 
 28691 Villanueva de la Ca\~{n}ada, Madrid, Spain
 \and
 Osservatorio Astrofisico di Arcetri, L.go E. Fermi 5, Firenze, Italy
 \and
 Department of Astronomy, Haverford College, Haverford, PA 19041, USA
 \and
Research and Scientific Support Department of ESA, ESTEC, Keplerlaan 1, 2201 AZ Noordwijk, The Netherlands
 \and
 Institute for Space Sciences, Bucharest-Magurele, Str. Atomostilor, 409, PoBox Mg-23,
Ro-077125, Romania
 \and
Lawrence Berkeley National Laboratory and Berkeley Center for Cosmological
Physics, Physics Department, University of California, Berkeley CA 94720, USA
 \and
 Department of Physics and Astronomy,
University of British Columbia, Vancouver, BC, V6T 1Z1 Canada
 \and
 University of Oxford, Astrophysics, Keble Road, Oxford, OX1 3RH, UK
 \and
Universit\'e Paris  7,  APC,  Case 7020, 75205 Paris Cedex 13, France
 \and
MilliLab, VTT Technical Research Centre of Finland, Information Technology
P.O. Box 1000, 02044 VTT, Finland
 \and
 Helsinki Institute of Physics,
P.O. Box 64, FIN-00014 Helsinki, Finland
\and
INAF-OABo, Istituto Nazionale di Astrofisica, Osservatorio Astronomico di Bologna, via Ranzani 1, I-40127 Bologna, Italy
 \and
 INFN, Istituto Nazionale di Fisica Nucleare, Sezione di Trieste, Via Valerio, 2, 34127, Trieste, Italy
\and
INFN, Istituto Nazionale di Fisica Nucleare, Sezione di ÒTor VergataÓ, Via della Ricerca Scientifica 1, I-00133 Roma, Italy
 \and
 University of California, Berkeley Space Sciences Lab 7 Gauss Way Berkeley, CA 94720, USA
\and
Computational Cosmology Center, Lawrence Berkeley National  Laboratory, Berkeley CA 94720, USA
\and
CESR, Centre d'Etude Spatiale des Rayonnements,
9, av du Colonel Roche, BP 44346 
31028 Toulouse Cedex 4, FRANCE
% ...
}


\abstract
{
This paper provides an overview of the Low Frequency Instrument (LFI)
programme within the ESA {\it Planck\/} mission.
The LFI instrument has been
developed to produce high precision maps of the microwave sky at
frequencies in the range 27--77$\,$GHz, below the peak of the cosmic
microwave background (CMB) radiation spectrum. 
The scientific goals 
are described, ranging from fundamental
cosmology to Galactic and extragalactic astrophysics. 
The instrument design and development are
outlined, together with the model philosophy and testing strategy. The
instrument is presented in the context of the {\it Planck\/} mission. The LFI
approach to ground and inflight calibration is described. We also
describe the LFI ground segment.
We present the results of a number of tests demonstrating the capability of
the LFI data processing centre (DPC) 
to properly reduce and analyse LFI flight data, from
telemetry information to calibrated and cleaned time ordered data, sky maps at each frequency (in temperature and polarization), component emission maps (CMB and diffuse foregrounds), catalogs for various classes of sources (the Early Release Compact Source Catalogue and the Final Compact Source Catalogue).
The organization 
of the LFI consortium is briefly presented as well as the role of the 
core team in data analysis and scientific exploitation.
All tests carried out on the LFI flight model demonstrate the excellent performance of the instrument and its various subunits.
The data analysis pipeline has been tested and its main steps verified.
In the first three months after launch,
the commissioning, calibration, performance, and verification phases
will be completed, after which {\it Planck\/} will begin its operational life,
in which LFI will have an integral part.
}
\authorrunning{Mandolesi et al.}
\titlerunning{The {\it Planck}-LFI programme}
\maketitle

\newpage

\section{Introduction}
\label{sec:intro}

\let\thefootnote\relax\footnotetext{The address to which the proofs
have to be sent is: \\
   Nazzareno Mandolesi\\
   INAF-IASF Bologna, Via Gobetti 101, I-40129, Bologna, Italy\\
   fax: +39-051-6398681\\
   e-mail: mandolesi@iasfbo.inaf.it}

In 1992, the Cosmic Background Explorer ({\it COBE\/}) team announced the 
discovery of intrinsic temperature fluctuations in the 
cosmic microwave background radiation (CMB; 
see Appendix \ref{acronyms} for a list of the acronyms appearing in this paper) 
on angular 
scales greater than $7^\circ$ and at a level of a few tens of $\mu$K
\citep{smoot_etal_92}.
One year later
two spaceborne CMB experiments were proposed to 
the European Space Agency (ESA)
in the framework of the Horizon 2000 Scientific Programme:  
the Cosmic Background Radiation Anisotropy Satellite \citep[COBRAS;][]{1994LNP...429..228M}, 
an array of receivers based on High Electron Mobility Transistor (HEMT) 
amplifiers; and the SAtellite for Measurement of Background Anisotropies
(SAMBA), an array of detectors based on bolometers \citep{1994ESAJ...18..239T}.  
The two proposals were accepted for an assessment 
study with the recommendation to merge.  In 1996, ESA selected a combined 
mission called COBRAS/SAMBA, subsequently renamed {\it Planck}, as the third 
Horizon 2000 Medium-Sized Mission.
Today {\it Planck\/} forms part of the ``Horizon 2000'' ESA Programme. 

The {\it Planck\/} CMB anisotropy
probe$^1$\footnote{$^1$ {\it Planck} \emph{(http://www.esa.int/Planck)} is a project of the European
Space Agency - ESA - with instruments provided by two scientific Consortia
funded by ESA member states (in particular the lead countries: 
France and Italy) with contributions from NASA (USA), and telescope
reflectors provided in a collaboration between ESA and a scientific
Consortium led and funded by Denmark.}, the first European and third generation mission
after {\it COBE\/} and {\it WMAP\/} (Wilkinson Microwave Anisotropy Probe), 
represents the state-of-the-art in precision cosmology today
\citep{Tauber_2009,bersa09,lamarreetal2009}.
The {\it Planck\/} payload (telescope instrument and cooling chain) is a
single, highly integrated
spaceborne CMB experiment. {\it Planck\/} is equipped with a $1.5$--m
effective aperture telescope
with two actively-cooled instruments that will scan the sky in nine frequency
channels from 
$30\,$GHz to $857\,$GHz: the Low Frequency Instrument (LFI) operating at
$20\,$K with pseudo-correlation radiometers, and the High Frequency Instrument
(HFI; \citealt{lamarreetal2009})
with bolometers operating at $100\,$mK. Each instrument has a specific role
in the programme.  The present paper describes the principal goals of LFI, its
instrument characteristics and programme.  The coordinated use of the two
different instrument technologies and analyses of their output data
will allow optimal control and suppression of systematic effects, including
discrimination of astrophysical sources. All the LFI channels and four of the
HFI channels will be sensitive to the linear polarisation of the CMB.
While HFI is more sensitive and should achieve higher angular resolution, 
the combination of the
two instruments is required to accurately subtract Galactic emission,
thereby allowing a reconstruction of the primordial CMB anisotropies
to high precision.

LFI (see \citealt{bersa09} for more details) consists of an array of 11 
corrugated horns 
feeding 22 polarisation-sensitive 
(see \citealt{paddy09} for more details)
pseudo-correlation radiometers 
based on HEMT transistors and 
MMIC technology, which are actively cooled to $20\,$K by 
a new concept sorption cooler specifically designed to deliver high 
efficiency, long duration cooling power
\citep{wade_bhandari_bowman,bhandari_prina_bowman,2009_LFI_SCS_T6}.
A differential scheme for the radiometers is adopted in which
the signal from the sky is compared with a stable reference load 
at $\sim4\,$K \citep{Valenziano2009}.
The radiometers cover three frequency bands centred on
30$\,$GHz, 44$\,$GHz, and 70$\,$GHz.
The design of the radiometers was
driven by the need to minimize the introduction of systematic errors and 
suppress noise fluctuations generated in the amplifiers.
Originally, LFI was to include seventeen $100$$\,$GHz horns with $34$ 
high sensitivity radiometers.
This system, which could have granted redundancy and cross-calibration with HFI as well as a cross-check of systematics, was not implemented.

The design of the horns is optimized to produce beams 
of the highest resolution in the sky and 
the lowest side lobes. 
Typical LFI main beams have full width half 
maximum (FWHM) resolutions of about $33'$, $27'$, 
and $13'$, respectively at 30$\,$GHz, 44$\,$GHz, and 70$\,$GHz, 
slightly superior to the requirements listed in Table \ref{table:reqs_sens}
for the cosmologically oriented $70$$\,$GHz channel. 
The beams are approximately elliptical with and ellipticity ratio
(i.e., major/minor axis) of $\simeq 1.15$--1.40.
The beam profiles will be measured in-flight by observing 
planets and strong radio sources
\citep{burigana_inflight_beamrec_2001}.

A summary of the LFI performance requirements adopted to help develop 
the instrument design is reported in 
Table~\ref{table:reqs_sens}.

\begin{table}[!ht]
  \caption{LFI performance requirements. The average sensitivity 
per $30^\prime$ pixel or per FWHM$^2$ resolution element 
($\delta$T and $\delta$T/T, respectively) is given in
CMB temperature units (i.e. equivalent thermodynamic temperature) for 14 
months of integration. The white noise (per frequency channel and 1~sec of 
integration) is given in antenna temperature units.
See Tables 2 and 6 for LFI measured performance.}
	\begin{tabular}{l c c c}
\hline
	Frequency channel &	$30\,$GHz	& $44\,$GHz	& $70\,$GHz \\
\hline
	InP detector technology	& MIC	& MIC	& MMIC \\
	Angular resolution [arcmin]	& 33 &	24 &	14 \\
	$\delta$T per $30^\prime$ pixel [$\mu$K] & 8	 & 8	& 8 \\
	$\delta$T/T per pixel [$\mu$K/K]	& 2.67    & 3.67 & 6.29 \\
	Number of radiometers (or feeds)	& 4 (2)	& 6 (3)	& 12 (6) \\
	Effective bandwidth [GHz]	& 6	&8.8	& 14 \\
	System noise temperature [K]	& 10.7	& 16.6	& 29.2 \\
	White noise per channel [$\mu$K $\cdot \sqrt{{\rm s}}$] & 116 & 113 & 105 \\
	Systematic effects [$\mu$K]	& $<$ 3	& $<$ 3	& $<$ 3 \\
\hline
\end{tabular}
\label{table:reqs_sens}
\end{table}

The constraints on the thermal behaviour, required to minimize systematic effects,
dictated a {\it Planck\/} cryogenic architecture that is one of the most complicated ever conceived for space.
Moreover, the spacecraft has been designed to exploit 
the favourable thermal conditions of the L2 orbit.
The thermal system is a combination of passive and active 
cooling:
passive radiators are used as thermal shields and pre-cooling stages,
while active cryocoolers are used both for instrument cooling
and pre-cooling.
The cryochain consists of the following main subsystems
\citep{1999_Collaudin}:

\begin{itemize}
    \item pre-cooling from $300\,$K to about $50\,$K by means of passive
radiators in three stages ($\sim150\,$K, $\sim100\,$K, $\sim50\,$K), 
which are called V-Grooves due to their conical shape;
    \item cooling to $18\,$K for LFI and pre-cooling the HFI $4\,$K cooler 
by means of a H$_2$ Joule-Thomson Cooler with sorption compressors
(the Sorption Cooler);
    \item cooling to $4\,$K to pre-cool the HFI dilution refrigerator 
and the LFI reference loads by means of a helium Joule-Thomson
cooler with mechanical compressors;
    \item cooling of the HFI to $1.6\,$K and finally $0.1\,$K with an open 
loop $^4$He--$^3$He dilution refrigerator.
\end{itemize}

The LFI front end unit is maintained at its operating 
temperature by the {\it Planck\/} H$_2$ Sorption Cooler Subsystem (SCS),
which is a closed-cycle vibration-free continuous cryocooler
designed to provide $1.2\,$W of cooling power at a temperature of $18\,$K.
Cooling is achieved by hydrogen compression,
expansion through a Joule-Thomson valve and liquid evaporation
at the cold stage. The {\it Planck\/} SCS is the first long-duration system 
of its kind to be flown on a space platform.
Operations and performance are described
in more detail in Sect.~\ref{sec:sorption} and 
in \cite{2009_LFI_SCS_T6}.

{\it Planck\/} is a spinning satellite. Thus, its receivers will observe the 
sky through a sequence of (almost great) circles 
following a 
scanning strategy (SS) aimed at 
minimizing systematic effects and achieving 
all-sky coverage for all receivers.
Several parameters are relevant to the SS.
The main one is the angle, $\alpha$, between the spacecraft spin axis 
and the telescope optical axis. Given the extension of the 
focal plane unit, each beam centre 
points to its specific angle, $\alpha_{\rm r}$.
The angle $\alpha$ is set to be $85^\circ$ to achieve a 
nearly all-sky coverage even in the so-called {\it nominal} SS 
in which the spacecraft spin axis is kept 
always exactly along the antisolar direction.
This choice avoids 
the ``degenerate'' case $\alpha_{\rm r} = 90^\circ$, 
characterized by a concentration of the crossings 
of scan circles only at the ecliptic poles 
and the consequent degradation of the quality of 
destriping and map-making codes 
\citep{burigana_etal_destr_97,maino_etal_destr_99,1996ApJ...458L..53W,1992issa.proc..391J}.
Since the {\it Planck\/} mission is designed
to minimize straylight contamination 
from the Sun, Earth, and Moon 
\citep{burigana_inflight_beamrec_2001,sandri_etal_2009_thisvolume},
it is possible
to introduce modulations of the spin axis
from the ecliptic plane to maximize the sky coverage,
keeping the solar aspect angle of the spacecraft constant
for thermal stability.
This drives us  towards the adopted {\it baseline} SS $^2$\footnote{$^2$ The above 
nominal SS is kept as a backup solution in case of a possible 
verification in-flight of unexpected problems with the {\it Planck\/} optics.} 
\citep{maris_etal_SS_2006}. 
Thus, the baseline SS adopts a cycloidal modulation of the 
spin axis, i.e. a precession around a nominal 
antisolar direction
with a semiamplitude cone of $7.5^\circ$. In this way,
all {\it Planck\/} receivers will cover the whole sky. 
A cycloidal modulation with a 6-month period satisfies the 
mission operational constraints,
while avoiding sharp gradients in the pixel hit count
\citep{dupac_tauber_2005}.
Furthermore, this solution allows one to spread the crossings of 
scan circles across a wide region that is beneficial to 
map-making, particularly for polarisation \citep{ashdown_etal_2007}. 
The last three SS parameters are: the sense
of precession (clockwise or anticlockwise);
the initial spin axis phase along the precession cone;
and, finally, the spacing between  
two consecutive spin axis repointings, chosen to be 
$2^\prime$ to achieve four all-sky surveys with the available guaranteed
number of spin axis manoeuvres.

Fifteen months of integration have been guaranteed since the approval of the mission. This will allow us to complete at least two all-sky surveys using all the receivers. The mission lifetime 
is going to be formally approved for an extension of 12 months, which will allow us to perform more than 4 complete sky surveys.

LFI is the result of an active collaboration between
about a hundred universities and research centres, in Europe, Canada, and 
USA, organized by the LFI consortium (supported by more than 300 
scientists) funded by national research and space agencies. 
The principal investigator leads a team of 
26 co-Investigators responsible for the development of the instrument 
hardware and software. The hardware was developed under the 
supervision of an instrument team. The data analysis and 
its scientific exploitation are mostly carried out by a core team,
working in close connection with the Data 
Processing Centre (DPC). 
The LFI core team is a diverse group of relevant 
scientists (currently $\sim$ 140) with the required expertise in instrument, data analysis, and theory to deliver to the wider {\it Planck\/} community the main mission data products. 
The core cosmology programme of Planck will be performed by the LFI and HFI core
teams.
The core team is closely linked to the wider {\it Planck\/} scientific community, 
consisting, besides the LFI consortium, of the HFI and Telescope consortia, which are
organized into various working groups.
{\it Planck\/} is managed by the ESA {\it Planck\/} science team.

The paper is organized as follows.
In Sect.~2, we describe the LFI cosmological and astrophysical objectives
and LFI's role in the overall mission.
We compare  the
LFI and {\it WMAP\/} sensitivities
with the CMB angular power spectrum in similar frequency bands, and 
discuss the cosmological improvement from {\it WMAP\/} represented by LFI
alone and in combination with HFI.
Section 3 describes the LFI optics, radiometers, and sorption cooler set-up
and performance.  The LFI programme is set forth in Sect.~4.
The LFI Data Processing Centre organisation is presented in Sect.~6, following
a report on the LFI tests and verifications in Sect.~5.
Our conclusions are presented in Sect.~7.

\section{Cosmology and astrophysics with LFI and {\it Planck}}
\label{sec:LFIscience}

{\it Planck\/} is the third generation space mission for CMB anisotropies
that will open a new era in our understanding of the Universe
\citep{bluebook}.
It will measure cosmological parameters with a 
much greater level of accuracy and precision than all previous efforts.
Furthermore, {\it Planck}'s high resolution all-sky survey,  
the first ever over this frequency range,
will provide a legacy to the astrophysical community for years to come.

\subsection{Cosmology}
\label{sec:cosmology}

The LFI instrument will play a crucial role for  cosmology.
Its LFI 70$\,$GHz channel is in a frequency window 
remarkably clear from foreground emission,
making it particularly advantageous for observing both CMB 
temperature and polarisation.
The two lower frequency channels at 30$\,$GHz and 44$\,$GHz
will accurately monitor Galactic and extra-Galactic foreground emissions 
(see Sect.~\ref{sec:astrophysics}), 
whose removal (see Sect.~\ref{sec:data_analysis})
is critical for a successful mission. 
This aspect is of key importance for CMB polarisation 
measurements since Galactic emission dominates the polarised sky.

The full exploitation of the cosmological information 
contained in the CMB maps will be largely based on the joint analysis of LFI 
and HFI data.
While a complete discussion of this aspect is beyond the scope of this paper,
in the next few subsections we discuss some topics of particular 
relevance to LFI or a combined analysis of LFI and HFI data.
In Sect.~\ref{CMB_APS_LFI}, we review the LFI sensitivity 
to the angular power spectrum on the basis of the realistic LFI 
sensitivity (see Table 6) and resolution (see Table 2) derived
from extensive tests.
This instrument description is adopted in Sect.~\ref{parameters} 
to estimate the LFI accuracy of the extraction of a representative set 
of cosmological parameters, alone and in combination with HFI.
Section~\ref{primoNG} addresses the problem of the detection of 
primordial non-Gaussianity, a topic of particular interest to the 
LFI consortium, which will require the combination 
of LFI and HFI, because of the necessity to clean the foreground.
On large signal angular scales, {\it WMAP\/} exhibits a minimum in the foreground
in the V band (61GHz, frequency range 53--69 GHz), 
thus we expect that the LFI 70$\,$GHz channel will be particularly 
helpful for investigating the CMB pattern on large scales, 
a topic discussed in Sect.~\ref{geometry_anomalies}. 

It is important to realise that these are just a few examples of what {\it Planck\/}
is capable of.  The increased sensitivity, fidelity and frequency range of
the maps, plus the dramatic improvement in polarisation capability will allow
a wide discovery space.  As well as measuring parameters, there will be tests
of inflationary models, consistency tests for dark energy models, and
significant new secondary science probes from correlations with other data-sets.

\begin{figure}
    \centering
    \includegraphics[width=0.7\hsize,angle=90]{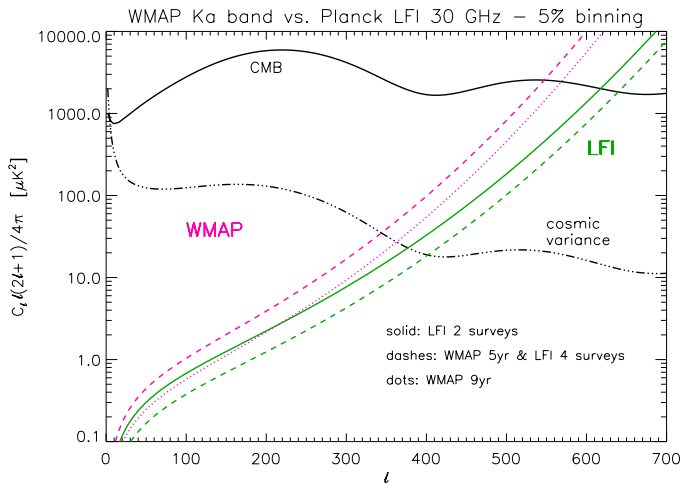}    
    \caption{CMB temperature anisotropy power spectrum (black solid line)
compatible with {\it WMAP\/} data is compared to {\it WMAP\/} (Ka band) and
LFI (30$\,$GHz) sensitivity, assuming subtraction of the
noise expectation, for different integration times as reported in the figure.
Two  {\it Planck\/} surveys correspond to about one year  of observations.
The plot shows separately the cosmic variance (black three dot-dashes) and the instrumental noise 
(red and green lines for {\it WMAP\/} and LFI, respectively) assuming a multipole binning of 5\%. 
This binning allows us to improve the sensitivity of the power spectrum estimation. 
For example, around $\ell =1000$ (100) this implies averaging the angular power spectrum over 50 (5) multipoles.
Regarding sampling variance, an all-sky survey is assumed here for simplicity.
    The use of the {\sc camb} code is acknowledged (see footnote $3$).}  
    \label{fig:aps_30_T}
\end{figure}

\begin{figure}
    \centering
    \includegraphics[width=0.7\hsize,angle=90]{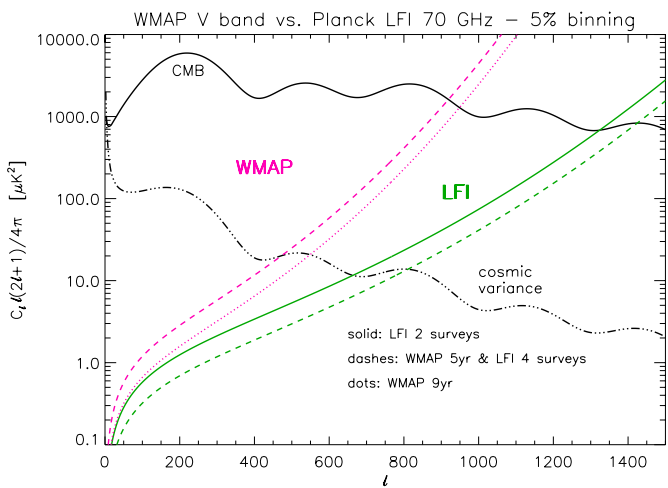}
    \caption{As in Fig.~\ref{fig:aps_30_T} but for the sensitivity of 
{\it WMAP\/} in V band and LFI at 70$\,$GHz.}
    \label{fig:aps_70_T}
\end{figure}

\subsubsection{Sensitivity to CMB angular power spectra}
\label{CMB_APS_LFI}

The statistical information encoded in CMB anisotropies, in both
temperature and polarisation, can be analyzed in terms of a 
``compressed'' estimator, the angular power spectrum, $C_\ell$
\citep[see e.g.,][]{ScottSmoot}.
Provided that the CMB anisotropies obey Gaussian statistics, as predicted in a 
wide class of models, the set of $C_\ell$s
contains most of the relevant statistical information.
The quality of the recovered power spectrum is a good predictor of the 
efficiency of extracting cosmological parameters
by comparing the theoretical 
predictions of Boltzmann codes$^3$ \footnote{$^3$ http://camb.info/}. 
Strictly speaking, this task must be carried out using 
likelihood analyses (see Sect.~\ref{sec:data_analysis}).
Neglecting systematic effects (and correlated noise), 
the sensitivity of a CMB anisotropy 
experiment to $C_\ell$, at each multipole $\ell$, is summarized by
the equation \citep{knox95}

\begin{equation}
\frac{\delta C_\ell}{C_\ell} \simeq \sqrt{\frac{2}{f_{\rm 
sky}(2\ell+1)}}\left[
1+\frac{A\sigma^2}{NC_\ell W_\ell}\right]\, ,
\label{fullvariance}
\end{equation}
where $A$ is the size of the surveyed area, 
$f_{\rm sky} = A / 4 \pi$,
$\sigma$ is the rms noise per pixel, 
$N$ is the total number of observed pixels, and $W_\ell$ is 
the beam window function. For a symmetric Gaussian beam,
$W_\ell = {\rm exp}(-\ell(\ell+1)\sigma_{\rm B}^2)$, where
$\sigma_{\rm B} = {\rm FWHM}/\sqrt{8{\rm ln}2}$
defines the beam resolution.

Even in the limit of an experiment of infinite sensitivity 
($\sigma=0$), the accuracy in the power spectrum is limited 
by so-called cosmic and sampling variance,
reducing to pure cosmic variance in the case of
all-sky coverage.
This dominates at low $\ell$ because of the 
relatively small number of available modes $m$ per multipole
in the spherical harmonic expansion of a sky map.
The multifrequency maps that will be obtained with {\it Planck\/}
will allow one to improve the foreground subtraction and 
maximize the effective sky area used in the 
analysis, thus improving our understanding of the CMB power spectrum
obtained from previous experiments.
However, the main benefits of the improved foreground subtraction will be in
terms of polarisation and non-Gaussianity tests.

Figures~\ref{fig:aps_30_T} and \ref{fig:aps_70_T} compare 
{\it WMAP\/}$^4$\footnote{$^4$ http://lambda.gsfc.nasa.gov/ } and LFI $^5$\footnote{$^5$ In this comparison, we exploit realistic LFI 
optical and instrumental performance as described in the following 
sections.} sensitivity to the CMB temperature $C_\ell$
at two similar frequency bands,
displaying separately the uncertainty originating in cosmic variance 
and instrumental performance and considering different project 
lifetimes.
For ease of comparison, we consider the same multipole binning
(in both cosmic variance and instrumental sensitivity).
The figures show how the multipole region, where cosmic 
variance dominates over instrumental sensitivity, 
moves to higher multipoles in the case of LFI and that the LFI 
70$\,$GHz channel allows us to extract information about an additional 
acoustic peak and two additional throats with respect to those achievable with the corresponding {\it WMAP\/} V band.

\begin{figure}
    \centering
    \includegraphics[width=0.75\hsize,angle=90]{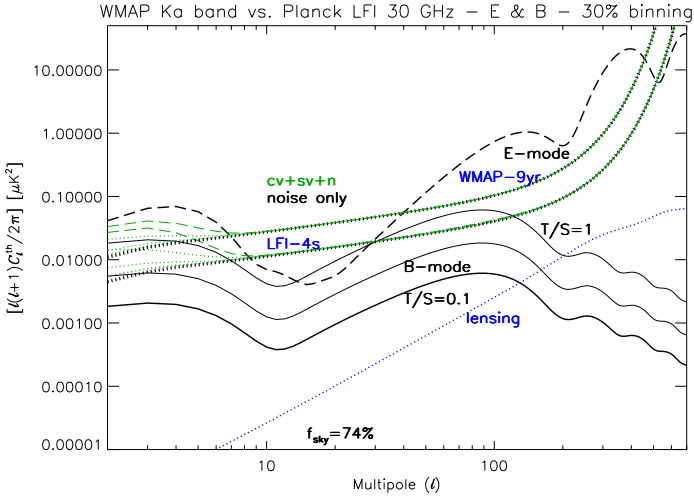}
    \caption{CMB E polarisation modes (black long dashes) compatible with {\it WMAP\/} data and CMB B polarisation modes (black solid lines) for different tensor-to-scalar ratios of primordial perturbations 
($r\equiv T/S = 1, 0.3, 0.1$, at increasing thickness)
are compared to {\it WMAP\/}
(Ka band, 9 years of observations) and LFI (30$\,$GHz, 4 surveys) 
sensitivity to the power spectrum,
assuming the noise expectation has been subtracted. 
The plots include cosmic and sampling variance plus instrumental noise 
(green dots for B modes, green long dashes for E modes, 
labeled with cv+sv+n; black thick dots, noise only) 
assuming a multipole binning of 30\% (see caption of Fig. 1 for the meaning of binning and of the number of sky surveys). 
Note that the cosmic and sampling (74\% sky coverage; as in WMAP polarization analysis, we exclude 
the sky regions mostly affected by Galactic emission) 
variance implies a dependence of the overall sensitivity 
at low multipoles on $r$ 
(again the green lines refer to 
$r = 1, 0.3, 0.1$, from top to bottom), 
which is relevant to the parameter estimation; 
instrumental noise only determines the capability of detecting the B mode. 
The B mode induced by lensing (blue dots) is also shown for comparison.}
    \label{fig:aps_30_P}
\end{figure}

\begin{figure}
\centering
    \includegraphics[width=0.75\hsize,angle=90]{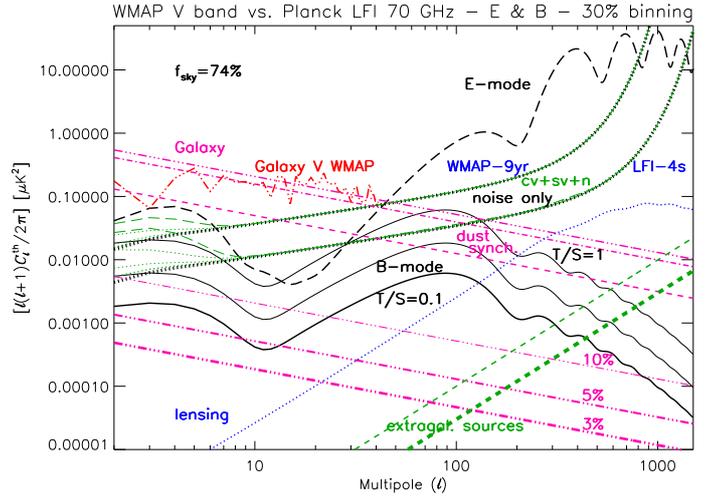}
    \caption{As in Fig.~\ref{fig:aps_30_P} but for the sensitivity of
{\it WMAP\/} in V band and LFI at 70$\,$GHz, and including also the comparison
with Galactic and extragalactic polarised foregrounds. Galactic synchrotron
(purple dashes) and dust (purple dot-dashes) polarised emissions produce the
overall Galactic foreground (purple three dot-dashes). {\it WMAP\/} 3-yr
power-law fits for uncorrelated dust and synchrotron have been used. For
comparison, {\it WMAP\/} 3-yr results derived directly from the foreground
maps using the HEALPix
package \citep{2005ApJ...622..759G}
are shown over a suitable multipole range: power-law
fits provide (generous) upper limits to the power at low multipoles.
Residual contamination levels by Galactic
foregrounds (purple three dot-dashes) are shown for 10\%, 5\%, and 3\% of the
map level, at increasing thickness. The residual
contribution of unsubtracted extragalactic sources, $C_\ell^{\rm res,PS}$ and
the corresponding uncertainty, $\delta C_\ell^{\rm res,PS}$
are also plotted as thick and thin green dashes.
These are computed assuming
a relative uncertainty $\delta \Pi / \Pi = \delta S_{\rm lim}/S_{\rm lim} =
10$\% in the knowledge of their degree of polarisation and the determination
of the source detection threshold.     
We assumed the same sky coverage as in Fig. 3. 
Clearly, foreground contamination is lower at 70 GHz than at 30 GHz, but, since CMB 
maps will be produced from a component separation layer (see 
Sects. 2.3 and 6.3) we considered the same sky region.}
    \label{fig:aps_70_P}
\end{figure}

As well as the temperature angular power spectrum, LFI can measure polarisation
anisotropies \citep{paddy09}.
A somewhat similar comparison is shown in Figs.~\ref{fig:aps_30_P} and 
\ref{fig:aps_70_P} but for the `E' and `B' polarisation modes,
considering in this case only the longest mission lifetimes
(9 yrs for {\it WMAP\/}, 4 surveys for {\it Planck\/})
reported in previous figures and a larger multipole binning; we note the 
increase in the signal-to-noise ratio compared to previous figures.
Clearly, foreground is more important for measurements of polarisation 
than for measurements of temperature. 
In the {\it WMAP\/} V band and the LFI 70$\,$GHz channels,
the polarised foreground is minimal (at least considering
a very large fraction of the sky and for the range of 
multipoles already explored by {\it WMAP\/}). Thus, we consider these optimal
frequencies to represent the potential uncertainty expected from
polarised foregrounds.  The Galactic foreground dominates over the 
CMB B mode and also the CMB E mode by up to multipoles of several tens.
However, foreground subtraction at an accuracy of 5$-$10\% of the map level 
is enough to reduce residual Galactic contamination to well below both
the CMB E mode and the CMB B mode for a wide range of multipoles
for $r = T/S \simeq 0.3$ (here $r$ is defined
in Fourier space).
If we are able to model Galactic polarised foregrounds with an accuracy at the
several percent level, then,
for the LFI 70$\,$GHz channel the main limitation
will come from instrumental noise. This will prevent an accurate
E mode evaluation at $\ell \sim 7$--20, or a B mode detection
for $r \lsim 0.3$. Clearly, a more accurate recovery of the polarisation modes 
will be possible from the exploitation of the {\it Planck\/} data at all frequencies.
In this context, LFI data will be crucial to model more accurately
the polarised synchrotron emission, which needs to be removed to greater than
the few percent level to detect
primordial B modes for $r \lsim 0.1$
\citep{2009arXiv0903.0345E}.

\subsubsection{Cosmological parameters}
\label{parameters}

Given the improvement relative to {\it WMAP\/} $C_\ell$ achievable 
with the higher sensitivity and resolution of {\it Planck\/}
(as discussed in the previous section for LFI), 
correspondingly superior determination of cosmological parameters is expected.
Of course, the better sensitivity and angular resolution of HFI channels
compared to {\it WMAP\/} and LFI ones will highly contribute to the
improvement in cosmological parameters
measured using {\it Planck}.

We present here the comparison between determinations of a suitable 
set of cosmological parameters using data from {\it WMAP}, {\it Planck},
and {\it Planck\/} LFI alone.

\begin{figure}
\centering
\includegraphics[width=1.\hsize]{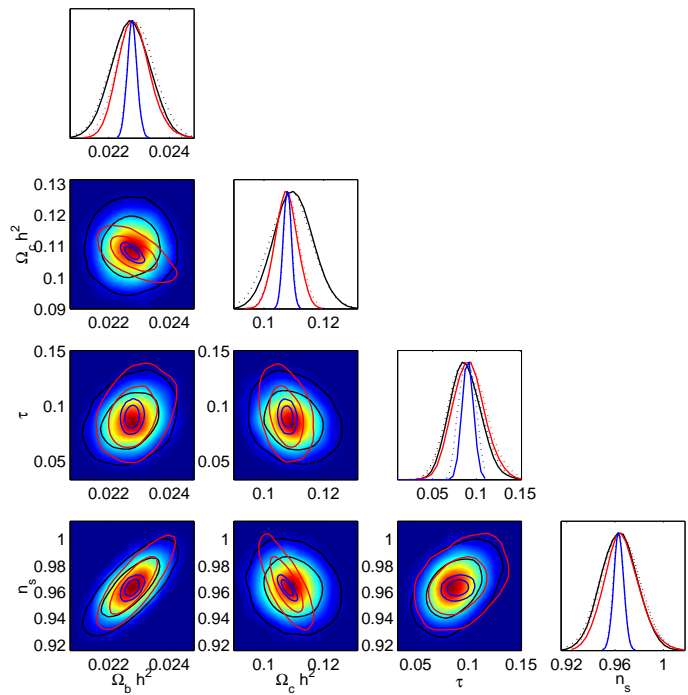}
\caption{Forecasts of 1$\sigma$ and 2$\sigma$ contours for the
cosmological parameters of the {\it WMAP\/}5 best-fit $\Lambda$CDM
cosmological model with reionization, as 
expected from {\it Planck\/} (blue lines) and from LFI alone (red lines) 
after 14 months of observations. 
The black contours are those obtained from {\it WMAP\/} five year observations. 
See the text for more details.}
\label{parametri}
\end{figure}

In Fig.~\ref{parametri}  we compare the forecasts for 1$\sigma$ and 
2$\sigma$ contours for 4 cosmological
parameters of the {\it WMAP\/}5 best-fit $\Lambda$CDM cosmological model:
the baryon density; the cold dark matter (CDM) density;
reionization, parametrized by the Thomson optical depth $\tau$;
and the slope of the initial power spectrum.  These results show the
expectation for
the {\it Planck\/} LFI 70$\,$GHz channel alone after 14 months of observations 
(red lines), 
the {\it Planck\/} combined 70$\,$GHz, 100$\,$GHz, and 143$\,$GHz 
channels for the same integration time 
(blue lines), and the {\it WMAP\/} five year observations (black lines). 
We assumed that the 70$\,$GHz channels and the 100$\,$GHz and 143$\,$GHz are the  
representative channels for LFI and HFI (we note that for HFI 
we have used angular resolution and sensitivities as given 
in Table 1.3
of the {\it Planck\/} scientific programme prepared by \citet{bluebook},) 
for cosmological purposes, 
respectively, and we assumed a coverage of $\sim70$\% of the sky. 
Figure ~\ref{parametri} shows that HFI $100\,$GHz and $143\,$GHz channels
are crucial for obtaining the most accurate cosmological parameter determination.

While we have not explicitly considered the other channels of 
LFI ($30\,$GHz and $44\,$GHz)
and HFI (at frequencies $\ge 217\,$GHz) we 
note that they are essential for achieving the accurate separation of the CMB from
astrophysical emissions, particularly for polarisation.

The improvement in cosmological parameter precision for LFI (2 surveys)
compared to {\it WMAP\/}5 \citep{Dunkley09,Komatsu09} is clear from Fig.~\ref{parametri}.
This is maximized 
for the dark matter abundance $\Omega_{\rm c}$ because of the 
performance of the LFI 70$\,$GHz channel with respect to {\it WMAP\/}5.
From Fig.~\ref{parametri}  it is clear that 
the expected improvement for {\it Planck\/} in cosmological parameter
determination 
compared to that of {\it WMAP\/}5 can open a new phase in our understanding
of cosmology.

\subsubsection{Primordial non-Gaussianity}
\label{primoNG}

Simple cosmological models assume Gaussian statistics for the anisotropies.
However, important information may come from mild deviations from Gaussianity
(see e.g., \citealt{BKMR04} for a review).
{\it Planck\/} total intensity and polarisation data will either provide the
first true measurement of non-Gaussianity (NG) in the primordial curvature
perturbations, 
or tighten the existing constraints (based on {\it WMAP\/} data, see footnote 3) by almost an
order of magnitude.

Probing primordial NG is another activity that requires foreground cleaned maps.
Hence, the full frequency maps of both instruments must be used for this
purpose.

It is very important that the primordial NG is 
{\em{model dependent}}. 
As a consequence of the assumed flatness of the inflaton potential, 
any intrinsic NG generated during standard single-field slow-roll 
inflation is generally small, hence   
adiabatic perturbations originated by quantum fluctuations of 
the inflaton field during standard inflation are nearly Gaussian distributed.  
Despite the simplicity of the inflationary paradigm, however, the 
mechanism by which perturbations are generated 
has not yet been fully established and various alternatives to the standard 
scenario have been considered. 
Non-standard scenarios for the generation of primordial perturbations
in single-field or multi-field inflation indeed permit higher NG levels. 
Alternative scenarios for the generation of the cosmological
perturbations, such as the so-called curvaton, the inhomogeneous 
reheating, and DBI scenarios \citep{2004PhRvD..70l3505A}, are characterized by a 
typically high NG level.
For this reason, detecting or even just constraining  
primordial NG signals in the CMB is one of the most promising 
ways to shed light on the physics of the early Universe. 

The standard way to parameterize primordial non-Gaussianity
involves the parameter $f_{\rm NL}$, which is typically small.
A positive detection of $f_{\rm NL} \sim 10$ would
imply that all standard single-field slow-roll models of inflation 
are ruled out. 
In contrast, an improvement to the limits on the amplitude of $f_{\rm NL}$ 
will allow one to strongly reduce the class of non-standard inflationary models 
allowed by the data, thus  
providing unique insight into the fluctuation generation mechanism. 
At the same time, {\it Planck\/} temperature and polarisation data will allow 
different predictions of the {\em shape} of non-Gaussianities to be tested
beyond the simple $f_{\rm NL}$ parameterization.   For simple, quadratic
non-Gaussianity of constant $f_{\rm NL}$, the angular bispectrum is
dominated by ``squeezed'' triangle configurations
with $\ell_1 \ll \ell_2,\ell_3$. 
This ``local'' NG is typical of models that
produce the perturbations immediately after inflation (such as for the curvaton 
or the inhomogeneous reheating scenarios). So-called DBI inflation models,
based on non-canonical kinetic terms for the {\em inflaton}
lead to non-local forms of NG, which are dominated by equilateral triangle
configurations.  It has been pointed out \citep{HT08} that 
excited initial states of the inflaton may lead to a third shape, 
called ``flattened'' triangle configuration.   

The strongest available CMB limits on $f_{\rm NL}$ for local NG comes from
{\it WMAP\/}5.  In particular, \cite{SSZ09}  obtained
$ -4 <  f_{\rm NL} < 80$ at 95\% confidence level (C.L.) using the optimal estimator of local NG. 
{\it Planck\/} total intensity and polarisation data will allow the window on
$|f_{\rm NL}|$ to be reduced below $\sim10$. \cite{2004PhRvD..70h3005B} 
and \cite{YKW07} demonstrated that a sensitivity to local non-Gaussianity
$\Delta f_{\rm NL} \approx 4$ (at $1\sigma$) is achievable with {\it Planck}.
We note that accurate measurement of E-type polarisation will play a significant
role in this constraint. 
Note also that the limits that {\it Planck\/} can achieve in this case are very
close to those of an ``ideal'' experiment. 
Equilateral-shape NG is less strongly constrained at present, with
$- 125 <  f_{\rm NL} <  435$ at 95\% C.L.  \citep{2009arXiv0905.3746S}. In this
case, {\it Planck\/} will also have a strong impact on
this constraint. Various authors \citep{SZ06,BR09}
have estimated that {\it Planck\/} data will allow us to reduce the bound 
on $\vert f_{\rm NL} \vert$ to around $70$.  

Measuring the primordial non-Gaussianity in CMB data to these levels of
precision  
requires accurate handling of possible contaminants, such as those 
introduced by instrumental noise and systematics, by the use of masks and
imperfect foreground and point source removal. 

\subsubsection{Large-scale anomalies}
\label{geometry_anomalies}

Observations of CMB anisotropies contributed significantly
to the development of the standard cosmological model,  
also known as the $\Lambda$CDM concordance model.
This involves a set of basic quantities for which CMB observations and other
cosmological and astrophysical data-sets agree: spatial curvature 
close to zero; $\simeq70$\% of the cosmic density in the form of dark energy;
$\simeq 20$\% in CDM; 
4$-$5\% in baryonic matter; and a nearly scale-invariant adiabatic, Gaussian
primordial perturbations.
Although the CMB anisotropy pattern obtained by 
{\it WMAP\/} is largely consistent with the concordance 
$\Lambda$CDM model, there are some interesting and 
curious deviations from it, in particular on the largest angular scales. 
Probing these deviations has required careful analysis procedures and so far
are at only modest levels of significance.  The anomalies can be listed as
follows:
\begin{itemize}
\item lack of power on large scales.
The angular correlation function is found to be uncorrelated 
(i.e., consistent with zero) for angles larger than $60^\circ$. 
In \citet{Copi07, Copi08}, it was shown that this event happens in only
0.03\% of realizations of the concordance model.  This is related to the
surprisingly low amplitude 
of the quadrupole term of the angular power spectrum
already found by {\it COBE\/}
\citep{smoot_etal_92,Hinshaw96}, and now confirmed by {\it WMAP\/}
\citep{Dunkley09,Komatsu09}. 
\item Hemispherical asymmetries. 
It is found that the power coming separately from the two hemispheres 
(defined by the ecliptic plane) is quite asymmetric, especially at low $\ell$
\citep{Eriksen04,Eriksen04Erratum,2004MNRAS.354..641H}.
\item Unlikely alignments of low multipoles. 
An unlikely (for a statistically isotropic random field) 
alignment of the quadrupole and the octupole 
\citep{Tegmark03,Copi04,Schwarz04,Weeks04,Land05}. 
Both quadrupole and octupole align 
with the CMB dipole \citep{Copi07}. 
Other unlikely alignments are described in
\citet{Abramo06}, \citet{2006PhRvL..96o1303W} and \citet{2007MNRAS.381..932V}. 
\item Cold Spot. 
\citet{Vielva04} detected a localized non-Gaussian behaviour in the southern
hemisphere using a wavelet analysis technique \citep[see also][]{Cruz05}.
\end{itemize}

It is still unknown whether these anomalies are indicative of new 
(and fundamental) physics beyond the concordance model 
or whether they are simply the residuals of imperfectly removed
astrophysical foreground or systematic effects. 
{\it Planck\/} data will provide a valuable contribution, not only in refining
the cosmological parameters of the standard cosmological model
but also in solving the aforementioned puzzles, because of the superior foreground 
removal and control of systematic effects, as well as {\it Planck}'s different
scan strategy and wider frequency range compared with {\it WMAP}.
In particular, the LFI 70$\,$GHz channel will be crucial,
since, as shown by {\it WMAP}, the foreground on large angular scales reaches a
minimum in the V band.

\subsection{Astrophysics}
\label{sec:astrophysics}

The accuracy of the extraction of the CMB anisotropy pattern from
{\it Planck\/} maps largely relies, particularly for polarisation, 
on the quality of the separation
of the {\it background} signal of cosmological origin from the various 
{\it foreground} sources of astrophysical origin that are
superimposed on the maps
(see also Sect.~\ref{sec:data_analysis}). 
The scientific case for {\it Planck\/} was presented by \cite{bluebook}
and foresees the full exploitation of the multifrequency data.  This is aimed 
not only at the extraction of the CMB, but also
at the separation and study of each astrophysical component,
using {\it Planck\/} data alone or in combination with other data-sets.
This section provides an update of the scientific case, with particular
emphasis on the contribution of the LFI to the science goals.

\subsubsection{Galactic astrophysics}
\label{sec:gal_astrophysics}

{\it Planck\/} will carry out an all-sky survey of the fluctuations in Galactic
emission at its nine frequency bands.
The HFI channels at $\nu \ge 100$$\,$GHz will provide the main improvement 
with respect to {\it COBE\/} characterizing the large-scale Galactic dust 
emission$^6$\footnote{$^6$ At far-IR frequencies
significantly higher than those covered by {\it Planck},
much information comes from {\it IRAS\/}
(see e.g., ~\citealt{iris} for a recent version of the maps).},
which is still poorly known, particularly in polarisation.
However, since Galactic dust emission 
still dominates over free-free and synchrotron at 70$\,$GHz 
(see e.g. \citep{wmap_fore_goldetal09} 
and references therein), LFI will provide crucial
information about the low frequency tail of this component.
The LFI frequency channels, 
in particular those at 30$\,$GHz and 44$\,$GHz,
will be relevant to the study of the diffuse, significantly polarised 
synchrotron emission and the almost unpolarised free-free 
emission.

Results from {\it WMAP\/}'s lowest frequency channels inferred an additional contribution, probably correlated 
with dust (see \citealt{Dobler2009} and references therein).
While a model with complex synchrotron emission 
pattern and spectral index cannot be excluded, several interpretations
of microwave \citep[see e.g.][]{hildebrandt_etal,bonaldi_etal_07}
and radio \citep{laporta_etal_2008} data, and in 
particular the ARCADE~2 results \citep{arcade2_kogut_etal_2009}, 
seem to support the identification 
of this anomalous component as spinning dust 
\citep{1998ApJ...494L..19D,Lazarian_Finkbeiner_2003}.
LFI data, at 30$\,$GHz in particular, will shed new light 
on this intriguing question.

Another interesting component that will be studied by {\it Planck\/}
data is the so-called ``haze'' emission in the inner Galactic region, 
possibly generated by synchrotron emission from relativistic electrons 
and positrons produced in the annihilations of dark matter particles 
(see e.g., \citealt{haze1,haze2,haze3} and references therein).
 
Furthermore, the full interpretation of the Galactic diffuse emissions 
in {\it Planck\/} maps will benefit from a joint analysis 
with both radio and far-IR data. 
For instance, PILOT \citep{pilot07}
will improve on Archeops results \citep{archeops_dust_05},
measuring polarised dust emission at frequencies higher than
353$\,$GHz, and BLAST-Pol \citep{BLAST-Pol}
at even higher frequencies. All-sky surveys 
at 1.4$\,$GHz (see e.g., \citealt{burigana_etal_2006} and references therein) 
and in the range of a few GHz to 15$\,$GHz will complement the low frequency 
side (see e.g., PGMS, \citealt{parkes_survey}; C-BASS, \citealt{cbass};
QUIJOTE, \citealt{Quijote}; and GEM, \citealt{gem}) allowing 
an accurate multifrequency analysis 
of the depolarisation phenomena at low and intermediate
Galactic latitudes. 
Detailed knowledge of the underlying noise properties in {\it Planck\/} maps 
will allow one to measure the correlation characteristics  
of the diffuse component, greatly 
improving physical models of the interstellar medium (ISM).
The ultimate goal of these studies is the development of a consistent
Galactic 3D model, which includes the various components of the ISM,
and large and small scale magnetic fields 
(see e.g., \citealt{hammurabi}),
and turbulence phenomena \citep{turbolence}.

While having moderate resolution and being limited in flux to a few hundred mJy, 
{\it Planck\/} will also provide multifrequency, all-sky information about
discrete Galactic sources.  This will include objects from the early stages of
massive stars to the late stages of stellar evolution \citep{latestages}, 
from HII regions to dust clouds \citep{dustclouds}.
Models for both the enrichment of the ISM and the interplay between stellar 
formation and ambient physical properties will be also tested. 

{\it Planck\/} will also have a chance to observe some Galactic micro-blazars
(such as e.g., Cygnus X-3) in a flare phase and perform multifrequency
monitoring of these events on timescales from hours to weeks.
A quick detection software (QDS) system was developed by a Finnish
group in collaboration with LFI DPC \citep{anne}.  This will be used
to identify
of source flux variation, in {\it Planck\/} time ordered data.

Finally, {\it Planck\/} will provide unique information for
modelling the emission from moving objects and diffuse interplanetary dust
in the Solar System. 
The mm and sub-mm emission from planets 
and up to 100 asteroids will also be studied \citep{asteroids,2009arXiv0902.0468M}.
The zodiacal light emission will also be measured 
to great accuracy, free from residual Galactic contamination \citep{ZLE}.

\subsubsection{Extragalactic astrophysics}
\label{sec:extrag_astrophysics}

The higher sensitivity and angular resolution of LFI compared to
{\it WMAP\/} will allow us to obtain substantially richer
samples of extragalactic sources at mm wavelengths. Applying a new
multi-frequency linear filtering technique to
realistic LFI simulations of the sky, \cite{2009MNRAS.394..510H} detected
1600, 1550, and 1000
sources with 95\% reliability at 30, 44, and 70$\,$GHz, respectively, over
about 85\% of the sky.  The 95\% completeness fluxes are 540, 340, and
270 mJy at 30, 44, and 70$\,$GHz, respectively.
For comparison, the total number of $|b|>5^\circ$ sources detected by
\cite{2009MNRAS.392..733M} at $\ge 5\sigma$ in {\it WMAP\/}5 maps at 33, 41,
and 61$\,$GHz (including several possibly spurious
objects), are 307, 301, and 161, respectively; the corresponding detection
limits increase from $\simeq 1\,$Jy at 23$\,$GHz,
to $\simeq 2\,$Jy at 61$\,$GHz. The number of detections reported by
\cite{2009ApJS..180..283W} is lower by about 20\%.

\begin{figure}
    \centering
    \includegraphics[width=0.75\hsize,angle=90]{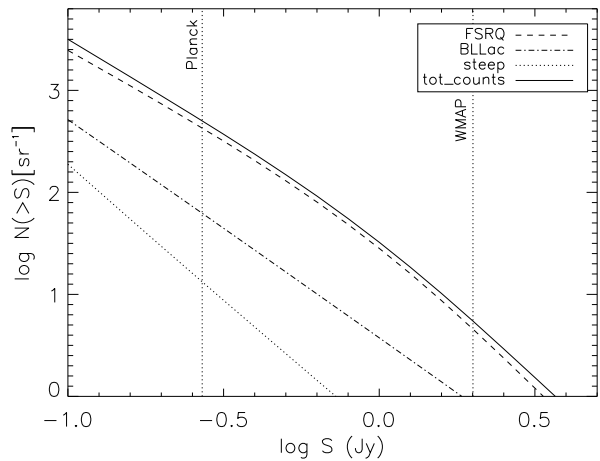}
    \caption{Integral counts of different radio source populations at 70$\,$GHz,
    predicted by the \cite{2005A&A...431..893D} model:
    flat-spectrum radio quasars; BL Lac objects; and steep-spectrum sources.
    The vertical dotted lines
    show the estimated completeness limits for {\it Planck\/} and {\it WMAP\/}
    (61$\,$GHz) surveys.}
    \label{fig:counts}
\end{figure}

As illustrated in Fig.~\ref{fig:counts}, the far larger source sample
expected from {\it Planck\/} will allow us to obtain good
statistics for different subpopulations of sources, some of which are not
(or only poorly) represented in the {\it WMAP\/} sample.
The dominant radio population at LFI frequencies consists of flat-spectrum
radio quasars, for which LFI will
provide a bright sample of $\ge 1000$ objects, well suited to cover the
parameter space of current physical models.
Interestingly, the expected numbers of blazars and BL Lac objects detectable by LFI are similar to those expected from
the {\it Fermi Gamma-ray Space Telescope\/} (formerly {\it GLAST\/};
\citealt{2009arXiv0902.1559A,2009arXiv0902.1089F}). It is
likely that the LFI and the Fermi blazar samples will have a substantial
overlap, making it possible to more carefully define 
the relationships between radio and gamma-ray properties of these sources than has been possible so far.
The analysis of spectral properties of the ATCA 20$\,$GHz bright sample
indicates that quite a few high-frequency selected
sources have peaked spectra; most of them are likely to be relatively old,
beamed objects (blazars), whose radio emission is
dominated by a single knot in the jet caught in a flaring phase.
The {\it Planck\/} sample will allow us to obtain key information
about the incidence and timescales of these flaring episodes, the distribution
of their peak frequencies, and therefore
the propagation of the flare along the jet. A small fraction of sources
exhibiting high frequency peaks may be extreme high
frequency peakers \citep{2000A&A...363..887D}, understood to be newly born radio
sources (ages as low as thousand years).
Obviously, the discovery of just a few of these sources would be extremely
important for sheding light on the poorly understood
mechanisms that trigger the radio activity of Galactic cores.

{\it WMAP\/} has detected polarised fluxes at $\ge 4\,\sigma$ in two or more
bands for only five extragalactic sources
\citep{2009ApJS..180..283W}. LFI will substantially improve on this, providing
polarisation measurements for tens of
sources, thus allowing us to obtain the first statistically meaningful unbiased
sample for polarisation studies at mm
wavelengths. It should be noted that {\it Planck\/} polarisation measurements will not be confusion-limited, as in the case of
total flux, but noise-limited. Thus the detection limit for polarised flux in {\it Planck}-LFI channels will be $\simeq
200$--$300\,$mJy, i.e., lower than for the total flux.

As mentioned above, the astrophysics programme of {\it Planck\/} is much
wider than that achievable with LFI alone, both because the
specific role of HFI and, in particular, the great scientific synergy
between the two instruments.  One noteworthy
example is the {\it Planck\/} contribution to the astrophysics of clusters.
{\it Planck\/} will detect $\approx 10^3$ galaxy
clusters out to redshifts of order unity by means of their thermal Sunyaev-Zel'dovich
effect \citep{2008A&A...491..597L,2008AN....329..147B}.
This sample will be extremely important for understanding both the formation
of large-scale structure and the physics of the intracluster medium.
To perform  these measurements, a broad spectral coverage,
i.e., the combination of data from both {\it Planck\/} instruments (LFI and HFI), is a key asset. This combination, supplemented
by ground-based, follow-up observations planned by the {\it Planck\/} team, will allow, in particular, accurate correction for
the contamination by radio sources (mostly due to the high quality of the LFI channels) and
dusty galaxies (HFI channels), either
associated with the clusters or in their foreground/background \citep{2009ApJ...694..992L}.

\subsection{Scientific data analysis}
\label{sec:data_analysis}

The data analysis process for a high precision experiment such as LFI must
be capable of reducing the data volume by several orders of magnitude
with minimal loss of information. The sheering size of the data set, 
the high sensitivity required to achieve the science goals,
and the significance
of the statistical and systematic sources of error all conspire to make data
analysis a far from trivial task.

The map-making layer provides a lossless compression by several orders of
magnitude, projecting the data set 
from the time domain to the discretized celestial sphere
\citep{1992issa.proc..391J,1994ApJ...436..452L,1996ApJ...458L..53W,
1997ApJ...480L..87T}.  Furthermore,  
timeline-specific instrumental effects that are not scan-synchronous are
reduced in magnitude when projected 
from time to pixel space  \citep[see e.g.,][]{2002A&A...384..736M} and, in
general, the analysis of maps provides a more convenient means of assessing the
level of systematics compared to timeline analysis.

Several map-making algorithms have been proposed to produce sky maps in 
total intensity (Stokes $I$) and linear polarisation (Stokes $Q$ and $U$) from 
the LFI timelines. So-called ``destriping'' algorithms have historically first
been applied. These take advantage of the details of the {\it Planck\/} scanning strategy to suppress correlated
noise \citep{maino_etal_destr_99}. Although computationally efficient, these methods do not, in general, yield 
a minimum variance map. To overcome this problem, minimum-variance map-making
algorithms have been devised and implemented specifically for LFI
\citep{2001A&A...372..346N,2005A&A...436.1159D}. The latter are also known as
generalized least squares (GLS) methods and are accurate and flexible. Their
drawback is that, at the size of the {\it Planck\/} data set, they require a significant amount of
massively powered computational resources
\citep{2006A&A...449.1311P,ashdown_etal_2007,ashdown_etal_2009} and are thus
infeasible to use within a Monte Carlo context. To overcome the limitations of
GLS algorithms, the LFI community has developed so-called ``hybrid'' algorithms
\citep{madam,madam-polar,madam-new}. These algorithms rely on a tunable
parameter connected to the $1/f$ knee frequency, a measure of the amount of
low frequency correlated noise in the time-ordered data: the higher the knee
frequency, the shorter the ``baseline'' length needed to be chosen to
properly suppress the $1/f$ contribution. From this point of view,
the GLS solution can
be thought of as the limiting case when the baseline length approaches the
sampling interval.
Provided that the knee frequency is not too high, hybrid algorithms can achieve
GLS accuracy at a fraction of the computational demand. Furthermore, they can
be tuned to the desired precision when speed is an issue (e.g., for
timeline-to-map Monte Carlo production). The baseline map-making algorithms
for LFI is a hybrid code dubbed {\sc madam}.

Map-making algorithms can, in general, compute the correlation (inverse
covariance) matrix of the map estimate that they produce \citep{ncvm}. 
At high resolution this computation, though feasible, is impractical, because the size of
the matrix makes its handling and inversion prohibitively difficult. 
At low resolution, the covariance matrix will be produced instead: this is of
extreme importance for the accurate characterization of the low multipoles of
the CMB \citep{ncvm,bolpol}.

A key tier of {\it Planck\/} data analysis is the separation of astrophysical
from cosmological components. A variety of methods have been developed to this
end \citep[e.g.,][]{2008A&A...491..597L}. Point source extraction 
is achieved by exploiting non-{\it Planck\/} catalogues, as well as filtering
{\it Planck\/} maps with optimal functions (wavelets) capable of recognizing
beam-like patterns. In addition to linearly 
combining the maps or fitting for known templates, diffuse emissions are
separated by using the statistical distributions of the different components,
assuming independence between them, 
or by means of a suitable parametrization and fitting of foreground unknowns on 
the basis of spatial correlations in the data or, in alternative,
multi-frequency single resolution elements only.

The extraction of statistical information from the CMB usually proceeds by means of 
correlation functions. Since the CMB field is Gaussian to a large extent
\citep[e.g.][]{SSZ09}, most of the information is encoded in the two-point
function or equivalently in its reciprocal representation in spherical
harmonics space. Assuming rotational invariance, the latter quantity is well
described by the set of $C_\ell$ \citep[see e.g.,][]{1994ApJ...430L..85G}.
For an ideal experiment, the estimated power spectrum could be directly
compared to a Boltzmann code prediction to constrain the cosmological
parameters. However, in the case of incomplete sky coverage (which induces
couplings among multipoles) and the presence of noise (which, in general, is
not rotationally invariant because of the coupling between correlated noise and
scanning strategy), a more thorough analysis is necessary. The likelihood
function for a Gaussian CMB sky can be easily written and provides a sound
mechanism for constraining models and data. The direct evaluation of this 
function, however, poses intractable computational issues. Fortunately, only
the lowest multipoles require exact treatment. This can be 
achieved either by direct evaluation in the pixel
domain or sampling the posterior distribution of the CMB using sampling
methods such as the Gibbs approach
\citep{2004ApJ...609....1J,2004PhRvD..70h3511W}.
At high multipoles, where the likelihood function cannot be evaluated exactly,
a wide range of effective, computationally affordable approximations exist
(see e.g., \citealt{2008PhRvD..77j3013H} 
and \citealt{grachaetal2009} and references therein). 
The low and high $\ell$ approaches to power spectrum estimation will be joined
into a hybrid procedure, pioneered by \cite{2004MNRAS.349..603E}.

The data analysis of LFI will require daunting computational resources.
In view of the size and complexity of its data set, accurate characterization
of the scientific results and error propagation will be achieved by means of a 
massive use of Monte Carlo simulations. A number of worldwide distributed
supercomputer centres will support the DPC in this activity. A partial list
includes NERSC-LBNL in the USA, CINECA in Italy, CSC in Finland,
and MARE NOSTRUM in Spain.
The European centres will benefit from the Distributed European Infrastructure
for Supercomputer Application $^7$\footnote{$^7$ http://www.deisa.eu}.

\section{Instrument}

\subsection{Optics}
\label{sec:LFIoptics}

During the design phase of LFI, great effort was dedicated to the 
optical design of the Focal Plane Unit (FPU).
As already mentioned in the introduction, the actual design of the
{\it Planck\/} 
telescope is derived from COBRAS and specially has been tuned by subsequent 
studies of the LFI team \citep{villa1998} and Thales-Alenia Space.
These studies demonstrated the importance of increasing the telescope diameter \citep{mandolesi2000}, optimizing the optical design, and also showed how complex it would be to match the real focal surface to the horn phase centres \citep{valenziano1998}. 
The optical design of LFI is the result of a long iteration process in 
which the optimization of the position and orientation of each feed horn involves a trade-off between angular resolution and sidelobe
rejection levels 
\citep{sandri_papI,burigana_papII,sandri_etal_2009_thisvolume}. 
Tight limits were also imposed by means of mechanical constraints. 
The 70$\,$GHz system has been improved in terms of the single horn design and its relative location in the focal surface.
As a result, the angular resolution has been maximized.

The feed horn development programme started
in the early stages of the mission with prototype 
demonstrators \citep{bersanelli1998}, followed by the elegant bread board 
\citep{villa2002} and finally by the qualification \citep{darcangelo2005}
and flight models (Villa et al.~2009). 
The horn design has a corrugated shape with a dual profile
\citep{gentili2000}. This choice was justified 
by the complexity of the optical interfaces (coupling with the telescope and focal plane horn accommodation)
and the need to respect the interfaces with HFI.

Each of the corrugated horns feeds an orthomode 
transducer (OMT) that splits the incoming signal into two orthogonal
polarised components \citep{2009_LFI_cal_O3}.
The polarisation capabilities of the LFI are guaranteed by the use of OMTs
placed immediately after the corrugated horns. 
While the incoming polarisation state is preserved inside the horn, the OMT
divides it into two linear orthogonal polarisations,
allowing LFI to measure the linear polarisation component of the incoming sky signal.  The typical 
value of OMT cross-polarisation is about $-30$dB, setting the spurious 
polarisation of the LFI optical interfaces at a level of $0.001$.

Table
2 shows the overall LFI optical 
characteristics expected in-flight \citep{Tauber_2009}.
The edge taper (ET) values, quoted in Table 2, refer to the horn taper;
they are reference values assumed during the design phase and do not
correspond to the true edge taper on the mirrors
(see \citealt{sandri_etal_2009_thisvolume} for details).
The reported angular resolution is the average full width half maximum (FWHM) of all the channels at the same frequency.
The cross-polar discrimination (XPD) is the ratio 
of the antenna solid angle of the cross-polar pattern to the 
antenna solid angle of the co-polar pattern, 
both calculated within the solid angle of the $-3\,$dB contour. 
The main- and sub-reflector spillovers 
represent the fraction of power that reach the horns without being 
intercepted by the main- and sub-reflectors, respectively. 

\begin{table}[!ht]
\label{table:opt1}
\begin{center}
 \caption{LFI optical performance. All the values are averaged over all
 channels at the same frequency. ET is the horn edge taper measured at
 $22^{\circ}$
 from the horn axis; 
 FWHM is the angular resolution in arcmin; $e$ is the ellipticity; XPD is the
cross-polar discrimination in dB; Ssp is the Sub-reflector spillover (\%); Msp
is the Main-reflector spillover (\%). See text for details.}
 
\begin{tabular}{l c c c c c c c}
\hline\hline
 &  ET & FWHM  & $e$  & XPD & Ssp & Msp \\
\hline
70 &$17\,$dB at $22^\circ$ &13.03&1.22&$-$34.73&0.17 & 0.65  \\
44 &$30\,$dB at $22^\circ$ &26.81&1.26&$-$30.54&0.074& 0.18 \\
30 &$30\,$dB at $22^\circ$&33.34&1.38&$-$32.37&0.24 & 0.59 \\
\hline
\end{tabular}
\end{center}
\end{table}

\subsection{Radiometers}
\label{sec:LFIinstrument}

\begin{figure}
    \centering
    \includegraphics[width=1.\hsize]{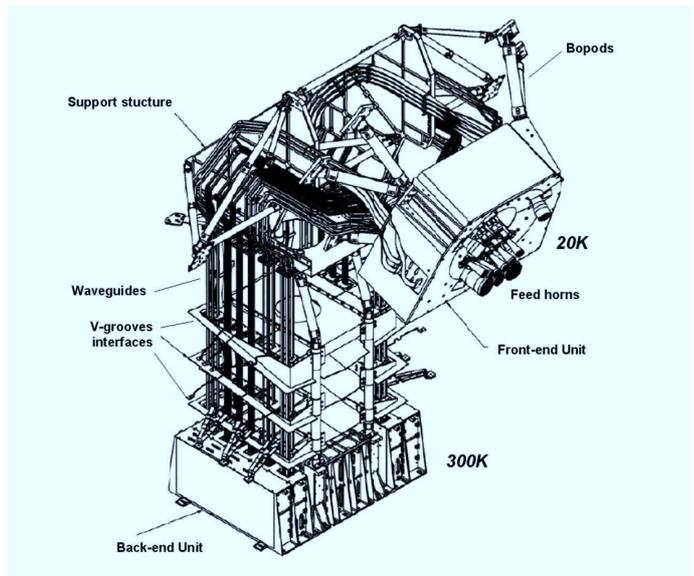}
    \caption{The LFI radiometer array assembly, with details of the
front-end and back-end units. The front-end radiometers are based on
wide-band low-noise amplifiers, fed by corrugated feedhorns which
collect the radiation from the telescope. A set of composite waveguides
transport the amplified signals from the front-end unit (at $20\,$K) to
the back-end unit (at $300\,$K). The waveguides are designed to meet
simultaneously radiometric, thermal, and mechanical requirements, and
are thermally linked to the three V-Groove thermal shields of the
{\it Planck\/} payload module. The back-end unit, located on top of the
{\it Planck\/}
service module, contains additional amplification as well as the
detectors, and is interfaced to the data acquisition electronics. The
HFI is inserted into and attached to the frame of the LFI focal-plane
unit.}
    \label{fig:Instrument_1}
\end{figure}

LFI is designed to cover the low frequency portion of the
wide-band {\it Planck\/} all-sky survey. A detailed description of the design
and implementation of the LFI instrument is given in 
\cite{bersa09}
and references therein, while the results of the on-ground
calibration and test campaign are presented in 
\cite{2009_LFI_cal_M3}
and
\cite{villaetal2009}.
The LFI is an array of cryogenically cooled
radiometers designed to observe in three frequency bands centered on
30$\,$GHz, 44$\,$GHz, and 70$\,$GHz with high sensitivity and practically no 
systematic error.
All channels are sensitive to the $I$, $Q$, and $U$ Stokes parameters,
thus providing information about both temperature and polarisation anisotropies.
The heart of the LFI instrument is a compact, 22-channel multifrequency
array of differential receivers with cryogenic low-noise amplifiers
based on indium phosphide (InP) high-electron-mobility transistors
(HEMTs). To minimise the power dissipation in the focal plane unit, which is
cooled to $20\,$K, the radiometers are divided into two subassemblies (the
front-end module, FEM, and the back-end module, BEM) connected by a set of
composite waveguides, as shown in Fig.~\ref{fig:Instrument_1}. Miniaturized,
low-loss passive components are implemented in the front end for optimal
performance and compatibility with the stringent thermo-mechanical
requirements of the interface with the HFI.

\begin{figure}
    \centering
    \includegraphics[width=1.\hsize]{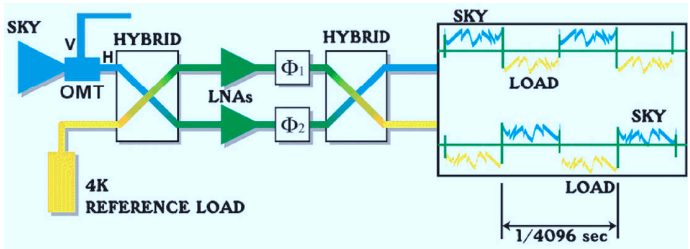}
    \caption{Schematic of the LFI front-end radiometer.
The front-end unit is located
at the focus of the {\it Planck\/} telescope, and comprises: dual-profiled
corrugated feed horns; low-loss ($0.2\,$dB), wideband ($> 20$\%) orthomode
transducers; and radiometer front-end modules with hybrids, cryogenic
low noise amplifiers, and phase switches. 
For details see \citet{bersa09}.}
    \label{fig:Instrument_2}
\end{figure}

The radiometer was designed to suppress $1/f$-type noise
induced by gain and noise temperature fluctuations in the amplifiers,
which would otherwise be unacceptably high for a simple, total-power system. A
differential pseudo-correlation scheme is adopted, in which signals from
the sky and from a black-body reference load are combined by a hybrid
coupler, amplified by two independent amplifier chains, and separated
by a second hybrid (Fig.~\ref{fig:Instrument_2}).
The sky and the reference load power
can then be measured and their difference calculated. Since the reference signal has
been affected by the same gain variations in the two amplifier chains as
the sky signal, the sky power can be recovered to high precision.
Insensitivity to fluctuations in the back-end amplifiers and detectors
is realized by switching phase shifters at $8\,$kHz synchronously in each
amplifier chain. The rejection of $1/f$ noise as well as immunity to
other systematic effects is optimised if the two input signals are
nearly equal. For this reason, the reference loads are cooled to $4\,$K 
\citep{Valenziano2009}
by mounting them on the $4\,$K structure of the HFI. In addition, the effect
of the residual offset ($< 1\,$K in nominal conditions) is reduced by
introducing a gain modulation factor in the onboard processing to
balance the output signal. As shown in Fig.~\ref{fig:Instrument_2},
the differencing receiver greatly improves the stability of the measured
signal (see also Fig.~8 in \citealt{bersa09}).

The LFI amplifiers at 30$\,$GHz and 44$\,$GHz use discrete InP HEMTs
incorporated into a microwave integrated circuit (MIC).
At these frequencies, the
parasitics and uncertainties introduced by the bond wires in a MIC
amplifier are controllable and the additional tuning flexibility
facilitates optimization for low noise. At 70$\,$GHz, there are twelve
detector chains. Amplifiers at these frequencies use monolithic
microwave integrated circuits (MMICs), which incorporate all circuit
elements and the HEMT transistors on a single InP chip. At these
frequencies, MMIC technology provides not only significantly superior
performance to MIC technology, but also allows faster assembly and
smaller sample-to-sample variance. Given the large number of amplifiers
required at 70$\,$GHz, MMIC technology can rightfully be regarded as an important
development for the LFI.

Fourty-four waveguides connect the LFI front-end unit, cooled to $20\,$K by
a hydrogen sorption cooler, to the back-end unit (BEU), which is mounted on
the top panel of the {\it Planck\/} service module (SVM) and maintained
at a temperature of
$300\,$K. The BEU comprises the eleven BEMs and the data acquisition
electronics (DAE) unit, which provides adjustable bias to the amplifiers
and phase switches as well as scientific signal conditioning. In the
back-end modules, the RF signals  are amplified further in the two
legs of the radiometers by room temperature amplifiers. The signals are
then filtered and detected by square-law detector diodes. A DC amplifier
then boosts the signal output, which is connected to the data acquisition
electronics. After onboard processing, provided by the radiometer box
electronics assembly (REBA), the compressed signals are down-linked to
the ground station together with housekeeping data. The sky and
reference load DC signals are transmitted to the ground as two separated
streams of data to ensure optimal calculation of the gain modulation
factor for minimal $1/f$ noise and systematic effects.
The complexity of the LFI system called for a highly modular plan of
testing and integration. Performance verification  was first carried out
at the single unit-level, followed by campaigns at sub-assembly and
instrument level, then completed with full functional tests after
integration into the {\it Planck\/} satellite. Scientific calibration has been
carried out in two main campaigns, first on the individual radiometer
chain assemblies (RCAs), i.e., the units comprising a feed horn and the
two pseudo-correlation radiometers connected to each arm of the
orthomode transducer (see Fig.~\ref{fig:Instrument_2}), and then at instrument level. For
the RCA campaign, we used sky loads and reference loads cooled close $4\,$K
which allowed us to perform an accurate verification of the instrument performance in
near-flight conditions. Instrument level tests were carried out with
loads at $20\,$K, which allowed us to verify the radiometer performance in
the integrated configuration. Testing at the RCA and instrument level, both for
the qualification model (QM) and the flight model (FM), were carried
out at Thales Alenia Space, Vimodrone (Milano, Italy). Finally,
system-level tests of the LFI integrated with HFI in the {\it Planck\/}
satellite were carried out at Centre Spatial de Liege (CSL)
in the summer of 2008.

\begin{figure}
    \centering
    \includegraphics[width=1.\hsize]{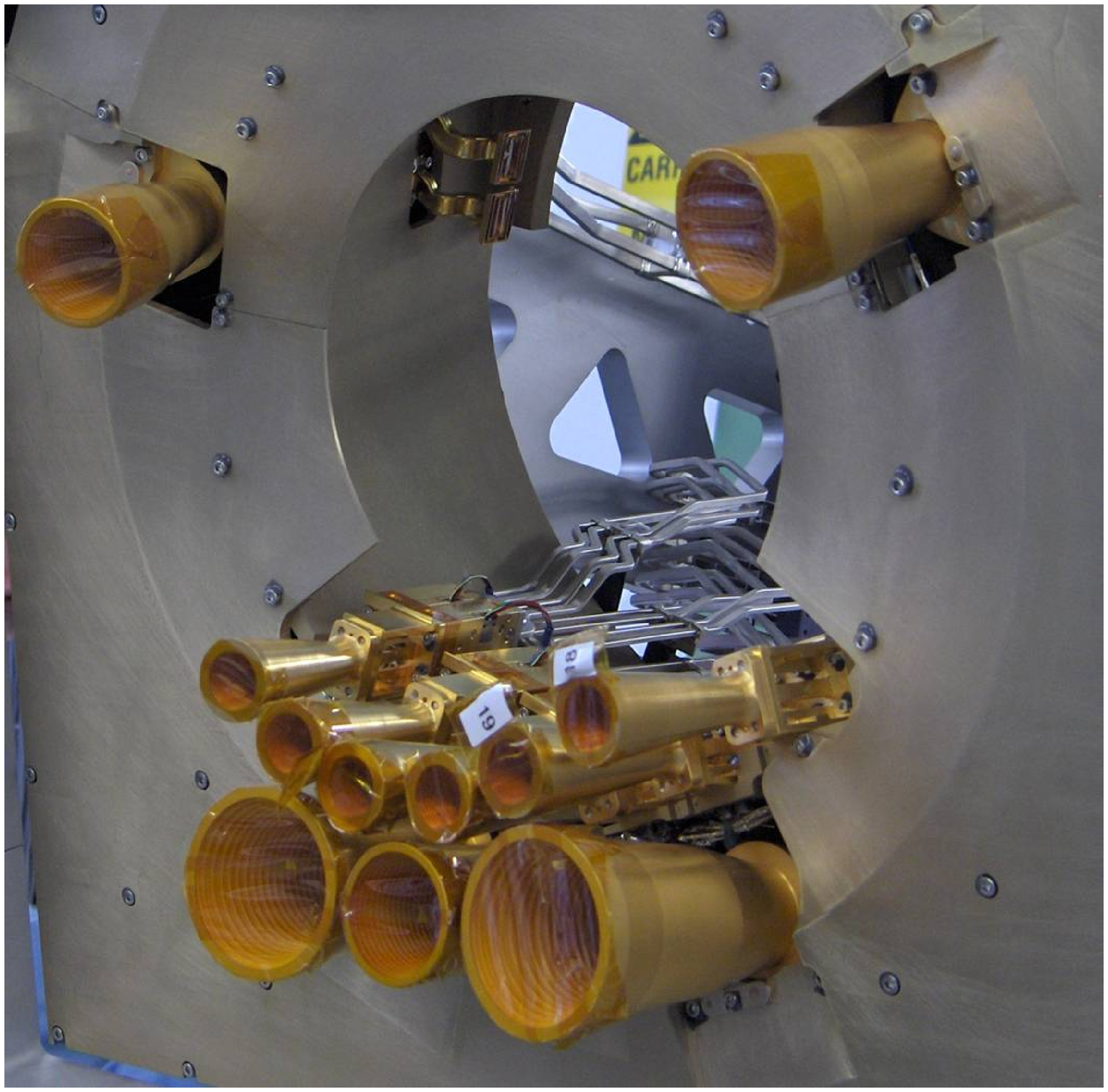}
    \includegraphics[width=1.\hsize]{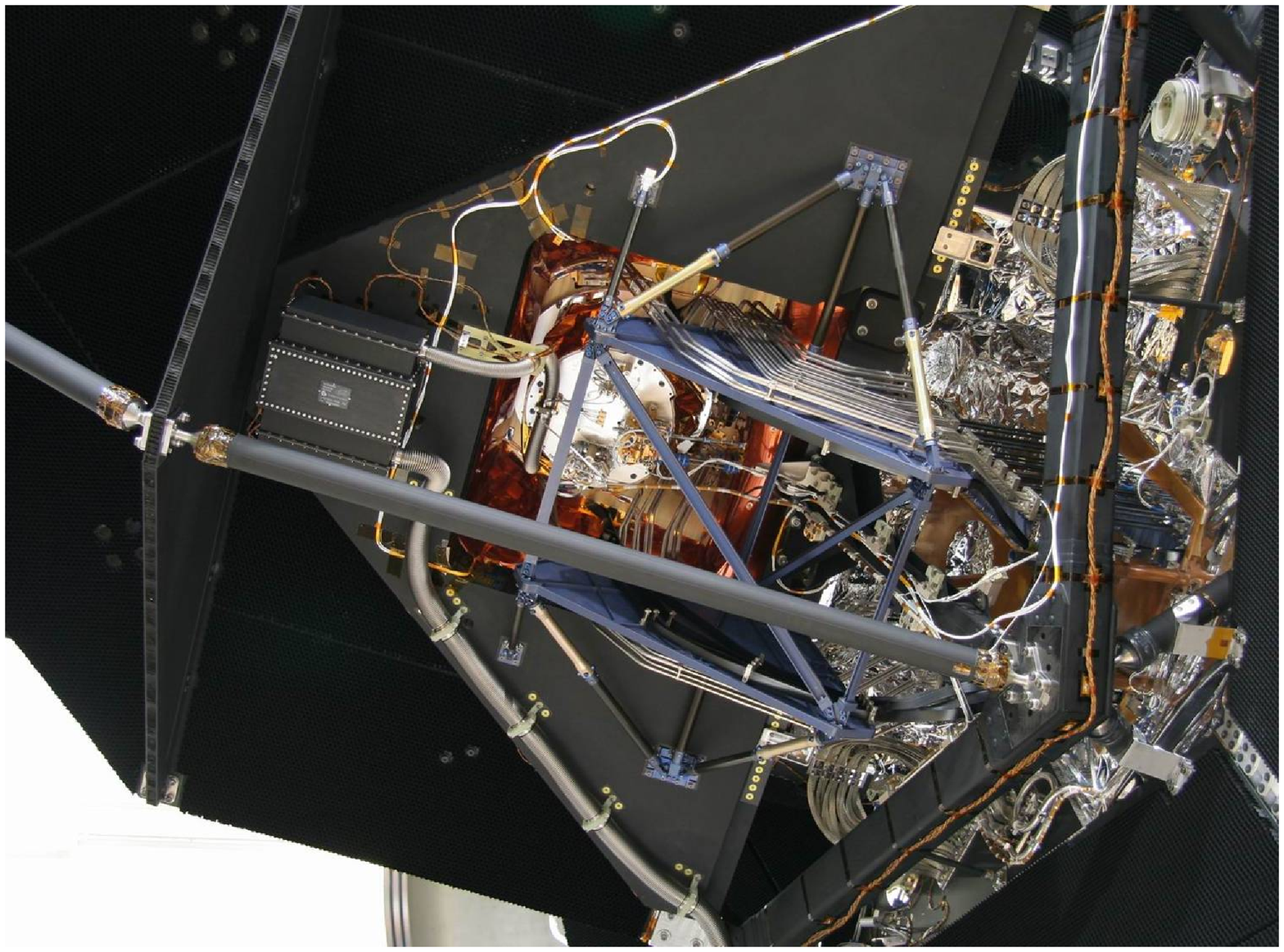}
    \caption{Top panel: picture of the LFI focal plane showing the feed-horns
and main frame. The central portion of the main frame is designed to
provide the interface to the HFI front-end unit, where the reference
loads for the LFI radiometers are located and cooled to 4K.
Bottom panel: A back-view of the LFI integrated on the {\it Planck\/} satellite.
Visible are the
upper sections of the waveguides interfacing the front-end unit, as well
as the mechanical support structure.}
    \label{fig:Instrument_3}
\end{figure}

\subsection{Sorption Cooler}
\label{sec:sorption}

The sorption cooler subsystem (SCS) is the first active element of the 
{\it Planck\/} cryochain. 
Its purpose is to cool the LFI radiometers to their operational 
temperature of around 20 K, while providing a pre-cooling stage for the HFI 
cooling system, a 4.5 K mechanical Joule-Thomson cooler and a Benoit-style 
open-cycle dilution refrigerator.
Two identical sorption coolers have been fabricated and assembled by 
the Jet Propulsion Laboratory (JPL) under contract to NASA. 
JPL has been a pioneer in the development and application of 
these cryo-coolers for space and the two {\it Planck\/} units are the first 
continuous closed-cycle hydrogen sorption coolers 
to be used for a space mission 
\citep{2009_LFI_SCS_T6}.

Sorption refrigerators are attractive systems for cooling instruments, 
detectors, and telescopes when a vibration-free system is required. 
Since pressurization and evacuation is accomplished by simply 
heating and cooling the sorbent elements sequentially, with no moving parts, 
they tend to be very robust and generate essentially no vibrations on the 
spacecraft. 
This provides excellent reliability and a long life. 
By cooling using Joule-Thomson (J-T) expansion through orifices, 
the cold end can also be 
located remotely (thermally and spatially) from the warm end. 
This allows excellent 
flexibility in integrating the cooler with the cold payload 
and the warm spacecraft.

\subsubsection{Specifications}
\label{sec:PlanckSCSSpecification}

The main requirements of the {\it Planck\/} SCS are summarized below:

\begin{itemize}
    \item Provision of about $1\,$W total heat lift at instrument interfaces 
using a $\leq60\,$K pre-cooling temperature at the coldest V-groove 
radiator on the {\it Planck\/} spacecraft;
    \item Maintain the following instrument interface temperatures:
    \begin{itemize}
    \item[] LFI at $\leq 22.5\,$K [80\% of total heat lift],
    \item[] HFI at $\leq 19.02\,$K [20\% of total heat lift];
    \end{itemize}
    \item Temperature stability (over one full cooler cycle $\approx 6000\,$s):
    \begin{itemize}
    \item[] $\leq 450\,$mK, peak-to-peak at HFI interface,
    \item[] $\leq 100\,$mK, peak-to-peak at LFI interface;
    \end{itemize}
    \item Input power consumption $\leq 470\,$W (at end of life, excluding electronics);
    \item Operational lifetime $\geq$ 2 years (including testing).
\end{itemize}

\subsubsection{Operations}
\label{sec:SorptionCoolerOperations}

\begin{figure}[!h]
    \centering
    \includegraphics[width=1.0\hsize]{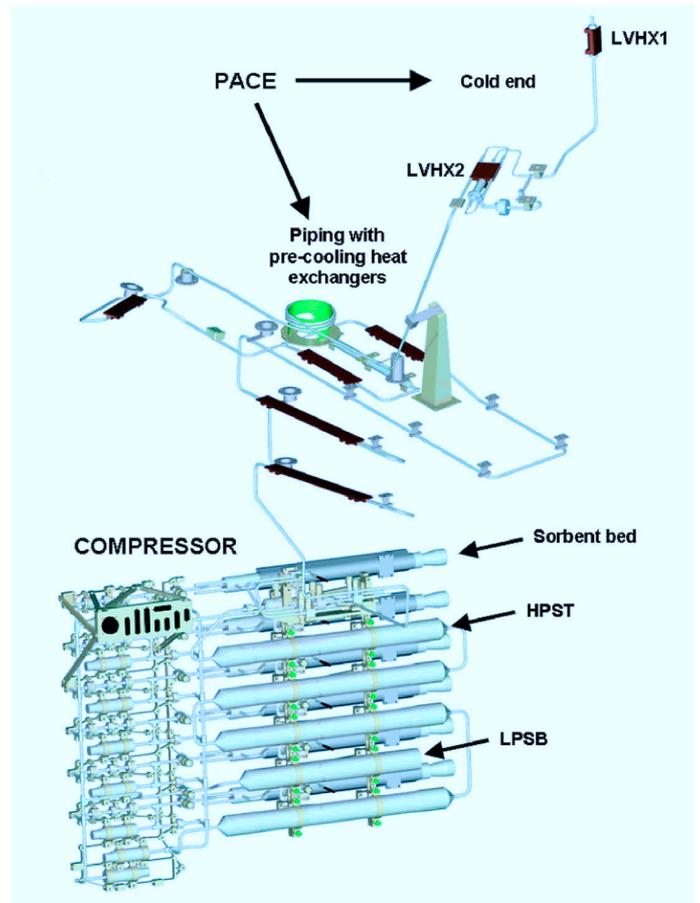}
    \caption{SCS thermo-mechanical unit. See Appendix~A for acronyms.}
    \label{fig:Figure_scs}
\end{figure}

The SCS consists of a thermo-mechanical unit 
(TMU, see Fig.~\ref{fig:Figure_scs}) and electronics to 
operate the system. 
Cooling is produced by J-T expansion with hydrogen as 
the working fluid. The key element of the $20\,$K sorption cooler is the 
compressor, an absorption machine that pumps hydrogen gas by thermally 
cycling six compressor elements (sorbent beds).
The principle of operation of the sorption compressor is based on the 
properties of a unique sorption material (a La, Ni, and Sn alloy), 
which can absorb a large amount of hydrogen at relatively low pressure, 
and desorb it to produce high-pressure gas when heated within a 
limited volume. Electrical resistances heat the 
sorbent, while cooling is achieved by thermally connecting, 
by means of gas-gap thermal switches, the 
compressor element to a warm radiator at $270\,$K on the satellite SVM. 
Each sorbent bed is connected to both the high-pressure and 
low-pressure sides of the plumbing system by check valves, 
which allow gas flow in a single direction only. 
To dampen oscillations on the high-pressure side of the compressor, 
a high-pressure stabilization tank (HPST) system is utilized. 
On the low-pressure side, a low-pressure storage bed (LPSB) 
filled with hydride, primarily operates as a storage bed for a large 
fraction of the H$_2$ inventory required to operate the cooler during 
flight and ground testing while minimizing the pressure in the 
non-operational cooler during launch and transportation. 
The compressor assembly mounts directly onto the warm radiator (WR) 
on the spacecraft.
Since each sorbent bed is taken through 
four steps (heat up, desorption, cool-down, absorption) in a cycle, 
it will intake low-pressure hydrogen and output high-pressure hydrogen 
on an intermittent basis. To produce a continuous stream of liquid 
refrigerant, the sorption beds phases are staggered so that at any given time, 
one is desorbing while the others are heating up, cooling down, 
or re-absorbing low-pressure gas. 

\begin{table*}[!ht]
    \centering
\begin{tabular}{cccccccc}
\hline
SCS Unit & Warm Rad. & 3 $^{\rm rd}$VGroove &
 \multicolumn{2}{c}{Cold-End $T$(K)}& Heat Lift & Input Power & Cycle Time \\
  & $T$(K)  & $T$(K)  & HFI I/F & LFI I/F  & (mW)  & (V)  & (s) \\
\hline
  & 270.5	& 45	& 17.2 & 18.7 $^{\mathrm{a,b}}$& 1100 & 297 &	940 \\
 Redundant	& 277	& 60	& 18.0 & 20.1 $^{\mathrm{a,b}}$& 1100 & 460
  & 492 \\
  & 282.6	& 60	& 18.4 & 19.9 $^{\mathrm{a,b}}$& 1050 & 388  & 667 \\
\hline
Nominal	& 270	& 47	& 17.1 & 18.7 $^{\mathrm{a}}$	& 1125 & 304	& 940 \\
    		& 273	& 48	& 17.5 & 18.7 $^{\mathrm{a}}$	&
 N/A	$^{\mathrm{c}}$& 470	& 525\\
\hline
\end{tabular}
\begin{list}{a}{}
\item[$^{\mathrm{a}}$] Measured at temperature stabilization assembly (TSA)
 stage
\item[$^{\mathrm{b}}$] In SCS-redundant test campaign TSA stage active control
 was not enabled 
\item[$^{\mathrm{c}}$] Not measured
\end{list}
    \label{tab:Table_scs}
  \caption{SCS flight units performance summary.}
\end{table*}

The compressed refrigerant then travels 
in the Piping and Cold-End Assembly (PACE, 
see Fig.~\ref{fig:Figure_scs}), through a 
series of heat exchangers linked to three V-Groove radiators on the 
spacecraft that provide passive cooling to approximately $50\,$K. 
Once pre-cooled to the 
required range of temperatures, the gas is expanded through the J-T valve. 
Upon expansion, hydrogen forms liquid droplets whose evaporation provides 
the cooling power. The liquid/vapour mixture then sequentially flows through 
the two Liquid Vapour Heat eXchangers (LVHXs) inside the cold end. 
LVHX1 and 2 are thermally and 
mechanically linked to the corresponding instrument (HFI and LFI) interface. 
The LFI is coupled to LVHX2 through an intermediate thermal stage, 
the temperature stabilization assembly (TSA). 
A feedback control loop (PID type), operated by the cooler electronics, 
is able to control the TSA peak-to-peak fluctuations down to the 
required level ($\leq100\,$mK).     
Heat from the instruments evaporates liquid hydrogen and the low pressure 
gaseous hydrogen is circulated back to the cold sorbent beds for 
compression.

\subsubsection{Performance}
\label{sec:PlanckSCSPerformance}

The two flight sorption cooler units were delivered to ESA in 2005. 
Prior to delivery, in early 2004, both flight models underwent 
subsystem-level thermal-vacuum test campaigns at JPL. 
In spring 2006 and summer 2008, respectively, SCS redundant and nominal
units were tested in cryogenic conditions on the spacecraft FM 
at the CSL facilities. 
The results of these two major test campaigns are summarized in 
Table~3
and reported in full detail in 
\cite{2009_LFI_SCS_T6}.

\section{LFI Programme}
\label{sec:programme}

The model philosophy adopted for LFI and the SCS was chosen to meet the
requirements of the ESA {\it Planck\/} system which assumed from the beginning
that there would be three development models of the satellite:

\begin{itemize}
\item The {\it Planck\/} avionics model (AVM) in which the system bus was shared
with the {\it Herschel\/} satellite, and allowed basic electrical interface
testing of all units and communications protocol and software interface
verification.
\item The {\it Planck\/} qualification model (QM), which was limited to the
{\it Planck\/} Payload Module (PPLM) containing QMs of LFI, HFI, and the {\it Planck\/}
telescope and structure that would allow a qualification vibration test
campaign to be performed at payload level, as well as alignment checks, and
would, in particular, allow a cryogenic qualification test campaign to be
performed on all the advanced instrumentation of the payload that had to
fully perform in cryogenic conditions.
\item The {\it Planck\/} protoflight model (PFM) which contained all the flight
model (FM) hardware and software that would undergo the PFM
environmental test campaign, culminating in extended thermal and
cryogenic functional performance tests.
\end{itemize}

\subsection{Model philosophy}
\label{LFI_Model_philosophy}

In correspondence with the system model philosophy, it was decided by the
{\it Planck\/} consortium to follow a conservative incremental approach
involving prototype demonstrators.

\subsubsection{Prototype demonstrators (PDs)}
\label{Prototype_demonstrators}

The scope of the PDs was to validate the LFI radiometer design concept
giving early results on intrinsic noise, particularly $1/f$ noise
properties, and characterise systematic effects in a preliminary fashion
to provide requirement inputs to the remainder of the instrument design and at
satellite level. The PDs also have the advantage of being able to test
and gain experience with very low noise HEMT amplifiers, hybrid
couplers, and phase switches. The PD development started early in the
programme during the ESA development pre-phase B activity and ran in
parallel with the successive instrument development phase of elegant
breadboarding.

\subsubsection{Elegant breadboarding (EBB)}
\label{Elegant_breadboarding}

The purpose of the LFI EBBs was to demonstrate the maturity of the full
radiometer design
across the whole frequency range of LFI prior to initiating qualification model construction. 
Thus, full comparison
radiometers (two channels covering a single polarisation
direction) were constructed, centred on 100$\,$GHz, 70$\,$GHz, and 30$\,$GHz,
extending from the expected design of the corrugated feed-horns at
their entrance to their output stages at their back-end.
These were put through functional and
performance tests with their front-end sections  operating at $20\,$K as
expected in-flight. It was towards the end of this development that the
financial difficulties that terminated the LFI 100$\,$GHz channel development
hit the programme.

\subsubsection{The qualification model (QM)}
\label{LFI_QM}

The development of the LFI QM commenced in parallel with the EBB
activities. From the very beginning, it was decided that only a limited
number of radiometer chain assemblies (RCA), each containing four
radiometers (and thus fully covering two orthogonal polarisation
directions) at each frequency should be included and that the remaining
instrumentation
would be represented by thermal mechanical dummies. Thus, the LFI QM
contained 2 RCA at 70$\,$GHz and one each at 44$\,$GHz and 30$\,$GHz. The active
components of the data acquisition electronics (DAE) were thus
dimensioned accordingly. The radiometer electronics box assembly (REBA)
QM supplied was a full unit. All units and assemblies went through
approved unit level qualification level testing prior to integration as
the LFI QM in the facilities of the instrument prime contractor Thales
Alenia Space Milano.

The financial difficulties also
disrupted the QM development and led to the use by ESA of a thermal-mechanical
representative dummy of LFI in the system level satellite QM test
campaign because of the ensuing delay in the availability of the LFI QM.
The LFI QM was however fundamental to the development of LFI as it enabled
the LFI consortium to perform representative
cryo-testing of a reduced model of the instrument and thus confirm the
design of the LFI flight model.

\subsubsection{The flight model (FM)}
\label{LFI_FM}

The LFI FM contained flight standard units and assemblies that went
through flight unit acceptance level tests prior to integration in to the
LFI FM. In addition, prior to mounting in the LFI FM, each RCA went
through a separate cryogenic test campaign after assembly to allow
preliminary tuning and confirm the overall
functional performance of each radiometer. At the LFI FM test level the
instrument went through an extended cryogenic test campaign that
included further tuning and instrument calibration that
could not be performed when mounted in the final configuration on the
satellite because of schedule and cost constraints. At the time of
delivery of the LFI FM to ESA for integration on the satellite, the only
significant verification test that remained to be done was the vibration
testing of the fully assembled radiometer array assembly (RAA).  This
could not be performed in a meaningful way at instrument level because of
the problem of simulating the coupled vibration input through the DAE
and the LFI FPU mounting to the RAA (and in particular into the
waveguides). Its verification was completed successfully during the
satellite PFM vibration test campaign.

\subsubsection{The avionics model (AVM)}
\label{LFI_AVM}

The LFI AVM was composed of the DAE QM, and its secondary power supply
box removed from the RAA of the LFI QM, an AVM model of the REBA and the
QM instrument harness. No radiometers were present in the LFI AVM, and
their active inputs on the DAE were terminated with resistors. The LFI
AVM was used successfully by ESA in the {\it Planck\/} System AVM test campaigns
to fulfill its scope outlined above.

\subsection{The sorption cooler subsystem (SCS) model philosophy}
\label{SCS_model_philosophy}

The SCS model development was designed to produce two coolers: a nominal
cooler and a redundant cooler. The early part of the model philosophy
adopted was similar to that of LFI, employing prototype development and
the testing of key components, such as single compressor beds, prior to the
building of an EBB containing a complete complement of components such as in
a cooler intended to fly. This EBB cooler was submitted to an intensive
functional and performance test campaign. The sorption cooler
electronics (SCE) meanwhile started development with an EBB and was
followed by a QM  and then FM1/FM2 build.

The TMUs of both the nominal and redundant sorption coolers went through
protoflight unit testing  prior to assembly with their respective PACE
for thermal/cryogenic testing before delivery. To conclude the
qualification of the PACE, a spare unit participated in the PPLM QM
system level vibration and cryogenic test campaign.

An important constraint in the ground operation of the sorption coolers
is that they could not be fully operated with their compressor beds far
from a horizontal position. This was to avoid permanent non-homogeneity
in the distribution of the hydrides in the compressor beds and the
ensuing loss in efficiency. In the fully integrated configuration of the
satellite (the PFM thermal and cryogenic test campaign) for test chamber
configuration, schedule and cost reasons would allow only one cooler to
be in a fully operable orientation. Thus, the first cooler to be
supplied, which was designated the redundant cooler (FM1), was mounted
with the PPLM QM and put through a cryogenic test campaign (termed PFM1)
with similar characteristics to those of the final thermal balance and
cryogenic tests of the fully integrated satellite. 
The FM1 was then later integrated into the satellite
where only short, fully powered, health checking was
performed. The second cooler was designated as the nominal cooler (FM2)
and participated fully in the final cryo-testing of the satellite. For
both coolers, final verification (TMU assembled with PACE) was achieved
during the {\it Planck\/} system-level vibration-test campaign and subsequent
tests.

The AVM of the SCS was supplied using the QM of the SCE and a simulator
of the TMU to simulate the power load of a real cooler.

\subsection{System level integration and test}
\label{System_level_integration_test}

The  {\it Planck\/} satellite and its instruments, were integrated at
the Thales Alenia Space facilities at Cannes in France.
The SCS nominal and redundant coolers were integrated onto the {\it Planck\/}
satellite before LFI and HFI.

Prior to integration on the satellite, the HFI FPU was integrated into
the FPU of LFI. This involved mounting the LFI $4\,$K loads onto HFI before
starting the main integration process, which was a very delicate
operation considering that when performed the closest approach of LFI and HFI
would be of the order of $2\,$mm. It should be remembered that LFI and HFI
had not ``met'' during the {\it Planck\/} QM activity and so this integration
was performed for the first time during the {\it Planck\/} PFM campaign. The
integration process had undergone much study and required special
rotatable ground support equipment (GSE) for the LFI RAA,
and a special suspension and balancing
system to allow HFI to be lifted and lowered into LFI at the correct
orientation along guide rails from above.  Fortunately the integration
was completed successfully.

Subsequently, the combined LFI RAA and HFI FPU were integrated onto the
satellite, supported by the LFI GSE, which was eventually removed during
integration to the telescope. The process of electrical integration and
checkout was then completed for LFI, the SCS and HFI, and the
proto-flight model test campaign commenced.

For LFI, this test campaign proceeded with ambient functional checkout
followed by detailed tests (as a complete subsystem prior to
participation with the SCS and HFI in the sequence of alignment),
electromagnetic compatibility (EMC),
sine and random acoustic vibration tests, and the sequence of system
level verification tests with the Mission Operations Control Centre (MOC,
at ESOC, Darmstadt) and LFI DPC. During all of these tests, at key points,
both the nominal and redundant SCS were put through ambient temperature
health checks to verify basic functionality.

The environmental test campaign culminated with the thermal balance and
cryogenic tests carried out at the Focal 5 facility of the Centre
Spatial de Liege, Belgium. The test was designed to follow very closely
the expected cool-down scenario after launch through to normal mission
operations, and it was during these tests that the two instruments and
the sorption cooler directly demonstrated together not only their
combined capabilities but also successfully met their operational margins.

\section{LFI test and verification}
\label{sec:test_verification}

\begin{figure*}[h!]
    \begin{center}
      \includegraphics[width=17.cm]{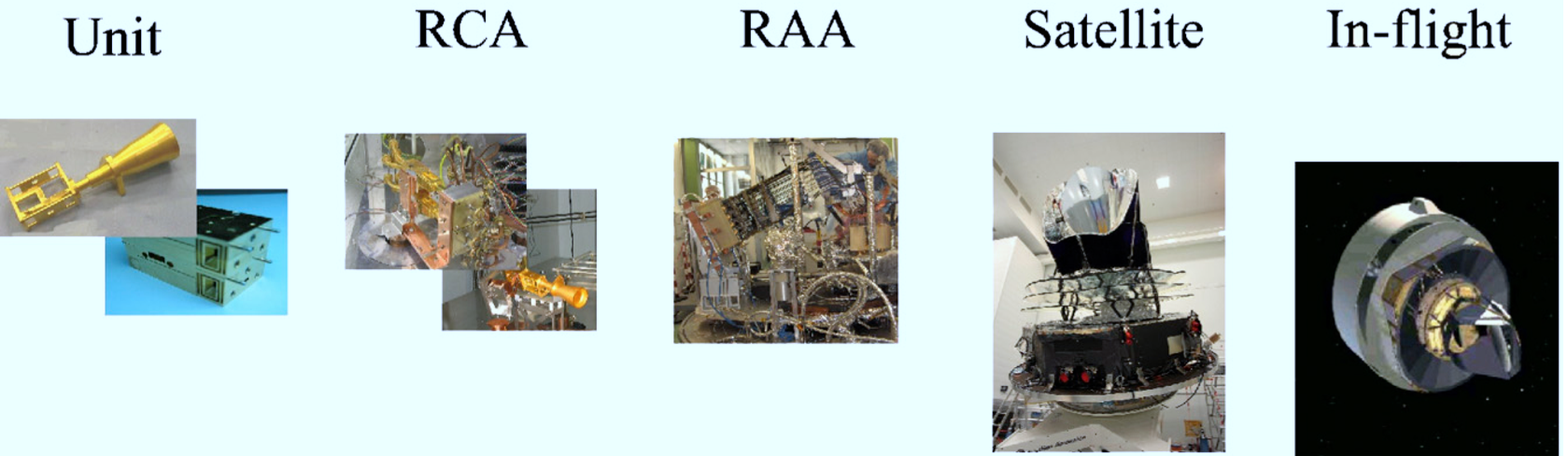} 
    \end{center}
  \caption{Schematic of the various calibration steps in the LFI 
development.}
  \label{fig:schematic_calibration_philosophy}
\end{figure*}

The LFI had been tested and calibrated before launch at various levels 
of integration, from the single components up to instrument and satellite 
levels; this approach, which is summarised schematically in 
Fig.~\ref{fig:schematic_calibration_philosophy}, provided inherent 
redundancy and optimal instrument knowledge. 

Passive components, i.e., feed-horns, OMTs, and waveguides, were tested 
at room conditions at the Plasma Physics Institute of the 
National Research Council (IFP-CNR) using a Vector Network Analyser. 
A summary of the measured performance parameters is provided 
in Table~\ref{tab:passive_components_parameters}; 
measurements and results are discussed in detail in 
\citet{2009_LFI_cal_O1} and \citet{2009_LFI_cal_O3, 2009_LFI_cal_O2}.

\begin{table}[h!]
    \caption{Measured performance parameters of the LFI passive 
components.}
    \label{tab:passive_components_parameters}
    \begin{center}
        \begin{tabular}{l p{6 cm}}
\vspace{.2cm}        
            \textbf{Feed Horns} & Return Loss $^1$, Cross-polar
($\pm$45$^\circ$) and Co-polar patterns (E, H and $\pm$45$^\circ$ planes) in
amplitude and phase, Edge taper at 22$^\circ$\\
\vspace{.2cm}        
            \textbf{OMTs} & Insertion Loss, Return Loss, Cross-polarisation,
 Isolation\\
        
            \textbf{Waveguides} & Insertion Loss, Return Loss, Isolation \\
        \end{tabular}
    \end{center}
    \footnotesize$^1$ Return loss and patterns (E,H for all frequencies, also
 $\pm 45^\circ$ and cross-polar for the $70\,$GHz system) have been measured
 for the assembly Feed Horn + OMT as well.
\end{table}

In addition, radiometric performance was measured several times during 
the LFI development on individual subunits 
(e.g., amplifiers, phase switches, detector diodes) 
on integrated front-end and back-end modules 
\citep{2009_LFI_cal_R8, 2009_LFI_cal_R9, 2009_LFI_cal_R10} 
and on the complete radiometric assemblies, both as 
independent RCAs \citep{villaetal2009} and in RAA, 
the final integrated instrument configuration \citep{2009_LFI_cal_M3}. 

In Table~\ref{tab:list_main_calibration_parameters} 
\citep[taken from][]{2009_LFI_cal_M3}, we list the main LFI 
radiometric performance parameters and the integration levels 
at which they have been measured. 
After the flight instrument test campaign, the LFI was cryogenically 
tested again after integration on the satellite with the HFI,
while the final characterisation will be performed in-flight 
before starting nominal operations.

\begin{table}[h!]
   \caption{Main calibration parameters and where they have been / will 
be measured. The following abbreviations have been used: 
SAT = Satellite; FLI = In-flight; FE = Front-end;
BE = Back-end; LNA = Low Noise Amplifier; PS = Phase Switch;
Radiom = Radiometric; and Susc = Susceptibility.}
   \label{tab:list_main_calibration_parameters}
   \begin{small}
   \begin{center}
      \begin{tabular}{|l|p{2cm}|c|c|c|c|}
         \hline
         \textbf{Category} & \textbf{Parameters} & \textbf{RCA} & \textbf{RAA} & \textbf{SAT} & \textbf{FLI} \\
         \hline
         \textbf{\textit{Tuning}} & FE LNAs &  Y & Y & Y & Y \\
               \cline{2-6}
                & FE PS & Y & Y & Y & Y \\
               \cline{2-6}
                & BE offset and gain & Y & Y & Y & Y \\
               \cline{2-6}
                & Quantisation / compression & N & Y & Y & Y \\
         \hline
         \textbf{\textit{Radiom.}}  & Photometric calibration & Y & Y & Y & Y \\
               \cline{2-6}
                & Linearity & Y & Y & Y & Y \\
               \cline{2-6}
                & Isolation & Y & Y & Y & Y \\
               \cline{2-6}
                & In-band response & Y & N & N & N \\
         \hline
         \textbf{\textit{Noise}}  & White noise & Y & Y & Y & Y \\
               \cline{2-6}
                & Knee freq.& Y & Y & Y & Y \\
               \cline{2-6}
                & 1/$f$ slope & Y & Y & Y & Y \\
         \hline
         \textbf{\textit{Susc.}} & FE temperature fluctuations & Y & Y & Y & Y \\
               \cline{2-6}
                        & BE temperature fluctuations & Y & Y & N & N \\
               \cline{2-6}
                        & FE bias fluctuations & Y & Y & N & N \\
         \hline
      \end{tabular}
   \end{center}
   \end{small}
\end{table}
                
The RCA and RAA test campaigns have been important to characterizing the instrument 
functionality and behaviour, and measuring its expected performance 
in flight conditions. In particular, 30$\,$GHz and 44$\,$GHz RCAs were 
integrated and tested in Italy, at the Thales Alenia Space (TAS-I) 
laboratories in Milan, while the 70$\,$GHz RCA test campaign was 
carried out in Finland at the Yilinen-Elektrobit laboratories 
\citep{villaetal2009}. After this testing phase, the 11~RCAs were 
collected and integrated with the flight electronics in the 
LFI main frame at the TAS-I labs, where the instrument final test 
and calibration has taken place \citep{2009_LFI_cal_M3}. 
Custom-designed cryofacilities \citep{2009_LFI_cal_T1,2009_LFI_cal_T2} 
and high-performance black-body input loads 
\citep{2009_LFI_cal_T4, 2009_LFI_cal_T5} were developed 
to test the LFI in the most flight-representative 
environmental conditions.

A particular point must be made about the front-end bias tuning,
which is a key step in determining the instrument scientific performance. 
Tight mass and power constraints called for a simple design of the 
DAE box so that power bias lines were divided into five 
common-grounded power groups with no bias voltage readouts. 
Only the total drain current flowing through the front-end amplifiers 
is measured and is available to the housekeeping telemetry.

This design has important implications for front-end bias tuning, 
which depends critically on the satellite electrical 
and thermal configuration. Therefore, this step was repeated 
at all integration stages and will also be repeated during ground 
satellite tests and in-flight before the start of nominal operations. 
Details about the bias tuning performed on front-end modules and 
on the individual integrated RCAs can be found in \citet{2009_LFI_cal_R8}, 
\citet{2009_LFI_cal_R10}, and \citet{villaetal2009}.

Parameters measured on the integrated instrument were found to be
essentially in line with measurements performed on individual receivers; 
in particular, the LFI shows excellent $1/f$ stability and rejection 
of instrumental systematic effects. On the other hand, the very ambitious 
sensitivity goals have not been fully met and the white noise sensitivity 
(see Table~\ref{table:meas_sens})
is $\sim$30\% higher than requirements. 
Nevertheless, the measured performance makes LFI the most sensitive 
instrument of its kind, a factor of 2 to 3 superior to 
{\it WMAP\/}$^8$\footnote{$^8$ Calculated on the final resolution element per unit 
integration time.} at the same frequencies.

\begin{table}[!ht]
  \caption{Calibrated white noise from ground-test results
extrapolated to the CMB input signal level. 
Two different methods are used to provide a reliable range of values 
(see \citealt{2009_LFI_cal_M3} for further details).
The final verification of sensitivity will be derived in-flight during 
the commissioning performance verification (CPV) phase.}
	\begin{tabular}{l c c c}
\hline
	Frequency channel &	$30\,$GHz	& $44\,$GHz	& $70\,$GHz \\
\hline
	White noise per $\nu$ channel & 141--154 & 152--160 & 130--146 \\

	$\;\;\;\;\;\;\;\;\;\;\;\;$[$\mu$K$\cdot\sqrt{{\rm s}}$] &      &    &  \\
\hline
\end{tabular}
\label{table:meas_sens}
\end{table}

\section{LFI Data Processing Centre (DPC)}
\label{sec:DPC}

To take maximum advantage of the capabilities of the {\it
Planck\/} mission and achieve its very ambitious scientific
objectives, proper data reduction and scientific analysis
procedures were defined, designed, and implemented very carefully.
The data processing was optimized so as to extract the maximum
amount of useful scientific information from the data set and
deliver the calibrated data to the broad scientific community
within a rather short period of time. As demonstrated by many
previous space missions using state-of-the-art technologies,
optimal scientific exploitation is obtained by combining the
robust, well-defined architecture of a data pipeline and its
associated tools with the high scientific creativity essential
when facing unpredictable features of the real data. Although many
steps required for the transformation of data were defined during
the development of the pipeline, since most of the foreseeable
ones have been implemented and tested during simulations, some of
them will remain unknown until flight data are obtained.

{\it Planck\/} is a PI mission, and its scientific achievements will
depend critically on the performance of the two instruments,
LFI and HFI, on the cooling chain, and on the telescope. The data
processing will be performed by two Data Processing Centres
\citep[DPCs,][]{Pas_2000_01, Pas_2000_02, Pas_2002}. However, despite the
existence of two separate distributed DPCs, the success of the
mission relies heavily on the combination of the measurements from
both instruments.

The development of the LFI DPC software has been performed in a
collaborative way across a consortium spread over 20 institutes in
a dozen countries. Individual scientists belonging to the software
prototyping team have developed prototype codes, which have then
been delivered to the LFI DPC integration team. The latter is
responsible for integrating, optimizing, and testing the code, and
has produced the pipeline software to be used during operations.
This development takes advantage of tools defined within the {\it
Planck\/} IDIS (integrated data and information system)
collaboration.

A software policy has defined, to allow the DPC perform the best
most superior algorithms within its pipeline, while fostering
collaboration inside the LFI consortium and across {\it Planck},
and preserving at the same time the intellectual property of the
code authors on the processing algorithms devised.

The {\it Planck\/} DPCs are responsible for the delivery and archiving of
the following scientific data products, which are the deliverables
of the {\it Planck\/} mission:
\begin{itemize}
    \item Calibrated time series data, for each receiver, after
    removal of systematic features and attitude reconstruction.
    \item Photometrically and astrometrically calibrated maps of
    the sky in each of the observed bands.
    \item Sky maps of the main astrophysical components.
    \item Catalogues of sources detected in the sky maps of the
    main astrophysical components.
    \item CMB power spectrum coefficients and an associated likelihood code.
\end{itemize}
Additional products, necessary for the total understanding of the
instrument, are being negotiated for inclusion in the {\it Planck\/}
Legacy Archive (PLA). The products foreseen to be added to the
formally defined products mentioned above are:
\begin{itemize}
    \item Data sets defining the estimated characteristics of
    each detector and the telescope (e.g. detectivity, emissivity,
    time response, main beam and side lobes, etc.).
    \item ``Internal'' data (e.g. calibration data-sets, data at
    intermediate level of processing).
    \item Ground calibration and assembly integration and verification
    (AIV) databases produced during
    the instrument development; and by gathering all information,
    data, and documents relative to the overall payload and all
    systems and subsystems. Most of this information is crucial
    for processing flight data and updating the knowledge and
    performance of the instrument.
\end{itemize}
The LFI DPC processing can be logically divided into three levels:
\begin{itemize}
    \item Level~1:
    includes monitoring of instrument health and behaviour and the definition
of corrective actions in the case of unsatisfactory function, and
the generation of time ordered information (TOI, a set of ordered
information on either a temporal or scan-phase basis), as well as
data display, checking, and analysis tools.
    \item Level~2: TOIs produced at Level~1 will be cleaned by removing
noise and many other types of systematic effects on the basis of
calibration information. The final product of the Level~2 includes
``frequency maps''.
    \item Level~3: ``Component maps'' will be generated by this level through
a decomposition of individual ``frequency maps'' and by also using
products from the other instrument and, possibly, ancillary data.
\end{itemize}
One additional level (``Level S'') is also implemented to develop
the most sophisticated simulations based on true instrument
parameters extracted during the ground test campaigns.

In the following sections, we describe the DPC Levels and the
software infrastructure, and we finally report briefly on the
tests that were applied to ensure that all pipelines are ready for
the launch.

\subsection{DPC Level~1}
\label{sec:LFIDPCLevel1}

Level~1 takes input from the MOC's data distribution system (DDS),
decompresses the raw data, and outputs time ordered information
for Level~2. Level~1 does not include scientific processing of the
data; actions are performed automatically by using pre-defined
input data and information from the technical teams. The inputs to
Level~1 are  telemetry (TM) and auxiliary data as they are
released by the MOC. Level~1 uses TM data to perform a routine
analysis (RTA -- real time assessment) of the spacecraft and
instrument status, in addition to what is performed at the MOC,
with the aim of monitoring the overall health of the payload and
detecting possible anomalies. A quick-look data analysis (TQL --
Telemetry Quick Look) of the science TM is also done, to monitor
the operation of the observation plan and verify the performance
of the instrument. This processing is meant to lead to the full
mission raw-data stream in a form suitable for subsequent data
processing by the DPC.

Level~1 also deals with all activities related to the production
of reports. This task includes the results of telemetry analysis,
but also the results of technical processing carried out on TOI to
understand the current and foreseen behaviour of the instrument.
This second item includes specific analysis of instrument
performance (LIFE -- LFI Integrated perFormance Evaluator), and
more general checking of time series (TSA -- Time Series Analysis)
for trend analysis purposes and comparison with the TOI from the
other instrument. The additional tasks of Level~1 relate to its
role as an instrument control and DPC interface with the MOC. In
particular, the following actions are performed:
\begin{itemize}
    \item Preparation of telecommanding procedures aimed at
    modifying the instrument setup.
    \item Preparation of Mission Information dataBases (MIBs).
    \item Communicate to the MOC ``longer-term'' inputs derived
    from feedback from DPC processing.
    \item Calibration of REBA parameters to fit long-term trends in the instrument setup.
\end{itemize}

\begin{figure}
    \centering
    \includegraphics[width=1.\hsize]{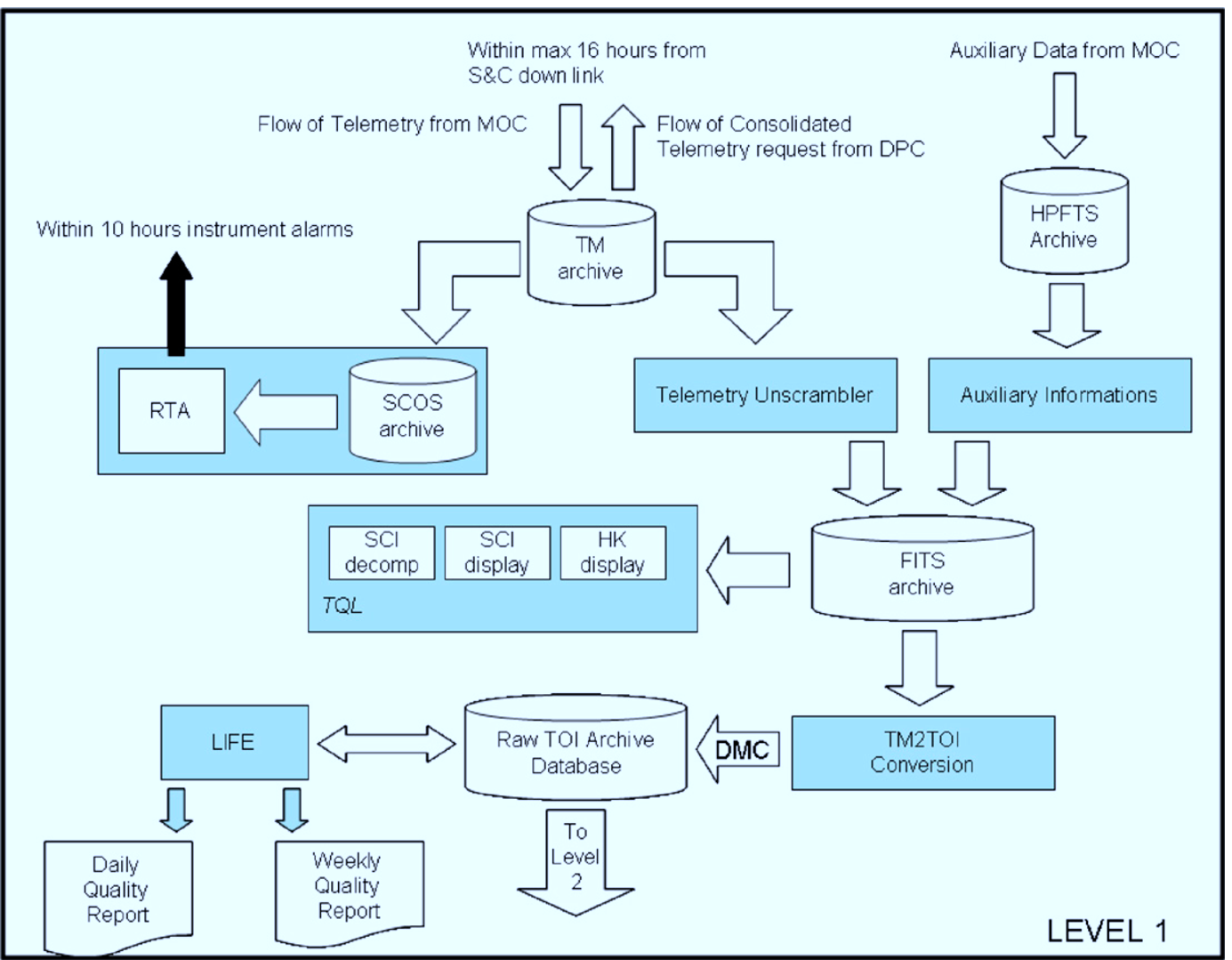}
    \caption{Level~1 structure.}
    \label{DPCL1}
\end{figure}

In Level~1, all actions are planned to be performed on a
``day-to-day'' basis during operation. In Fig.~\ref{DPCL1}, the
structure of Level~1 and required timings are shown. For more
details, we refer to \citet{Zac09}.

\subsection{DPC Level~2}
\label{sec:LFIDPCLevel2}

At this level, data processing steps requiring detailed instrument
knowledge (data reduction proper) will be performed. The raw time
series from Level~1 will also be used to reconstruct a number of
calibrated scans for each detector, as well as instrumental
performance and properties, and maps of the sky for each channel.
This processing is iterative, since simultaneous evaluation of
quite a number of parameters should be made before the
astrophysical signal can be isolated and averaged over all
detectors in each frequency channel. Continuous exchange of
information between the two DPCs will be necessary at Level~2 to
identify any suspect or unidentified behaviour or any results from
the detectors.

The first task that the Level~2 performs is the creation of
differenced data. Level~1 stores data from both Sky and Load.
These two have to be properly combined to produce differenced
data, therefore reducing the impact of $1/f$ noise achieved by
computing the so-called gain modulation factor $R$, which is
derived by taking the ratio of the mean signals from both Sky and
Load.

After differenced data are produced, the next step is the
photometric calibration that transforms the digital units into
physical units. This operation is quite complex: different methods
are implemented in the Level~2 pipeline that use the CMB dipole as
an absolute calibrator allowing for the conversion into physical
units.

Another major task is beam reconstruction, which is implemented
using information from planet crossings. An algorithm was
developed that performs a bi-variate approximation of the main
beam section of the antenna pattern and reconstructs the position
of the horn in the focal plane and its orientation with respect to
a reference axis.

The step following the production of calibrated timelines is the
creation of calibrated frequency maps. To achieve this, pointing
information will be encoded into time-ordered pixels i.e, ~pixel
numbers in the given pixelisation scheme (HEALPix) by identifying
a given pointing direction that is ordered in time. To produce
temperature maps, it is necessary to reconstruct the beam pattern
along the two polarisation directions for the main, intermediate,
and far parts of the beam pattern. This will allow the combination
of the two orthogonal components into a single temperature
timeline. On this temperature timeline, a map-making algorithm
will be applied to produce a map from each receiver.

\begin{figure*}
    \centering
    \includegraphics[width=1.\hsize]{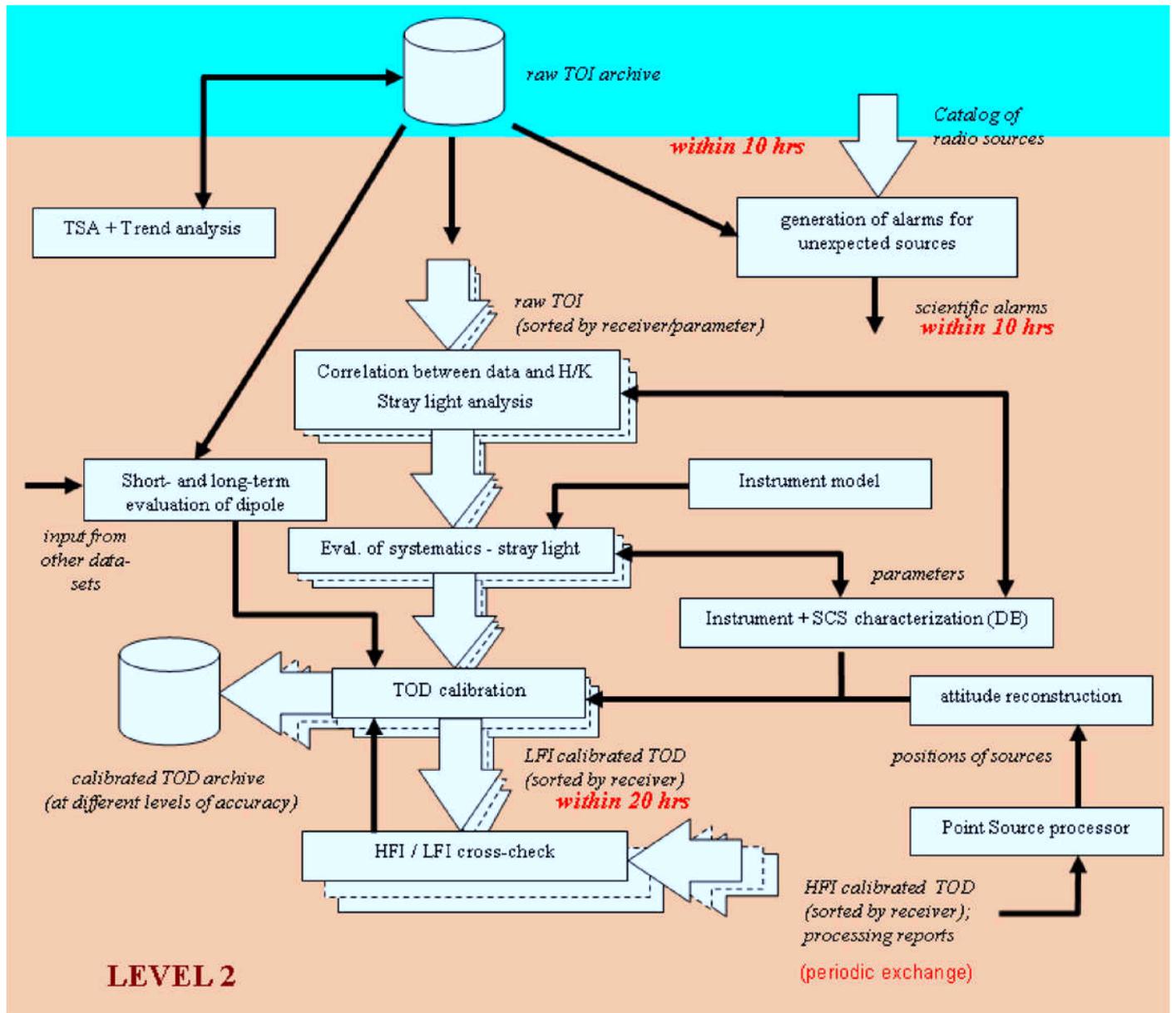}
    \caption{Level~2 calibration pipeline.}
    \label{Calibration}
\end{figure*}

\begin{figure*}
    \centering
    \includegraphics[width=1.\hsize]{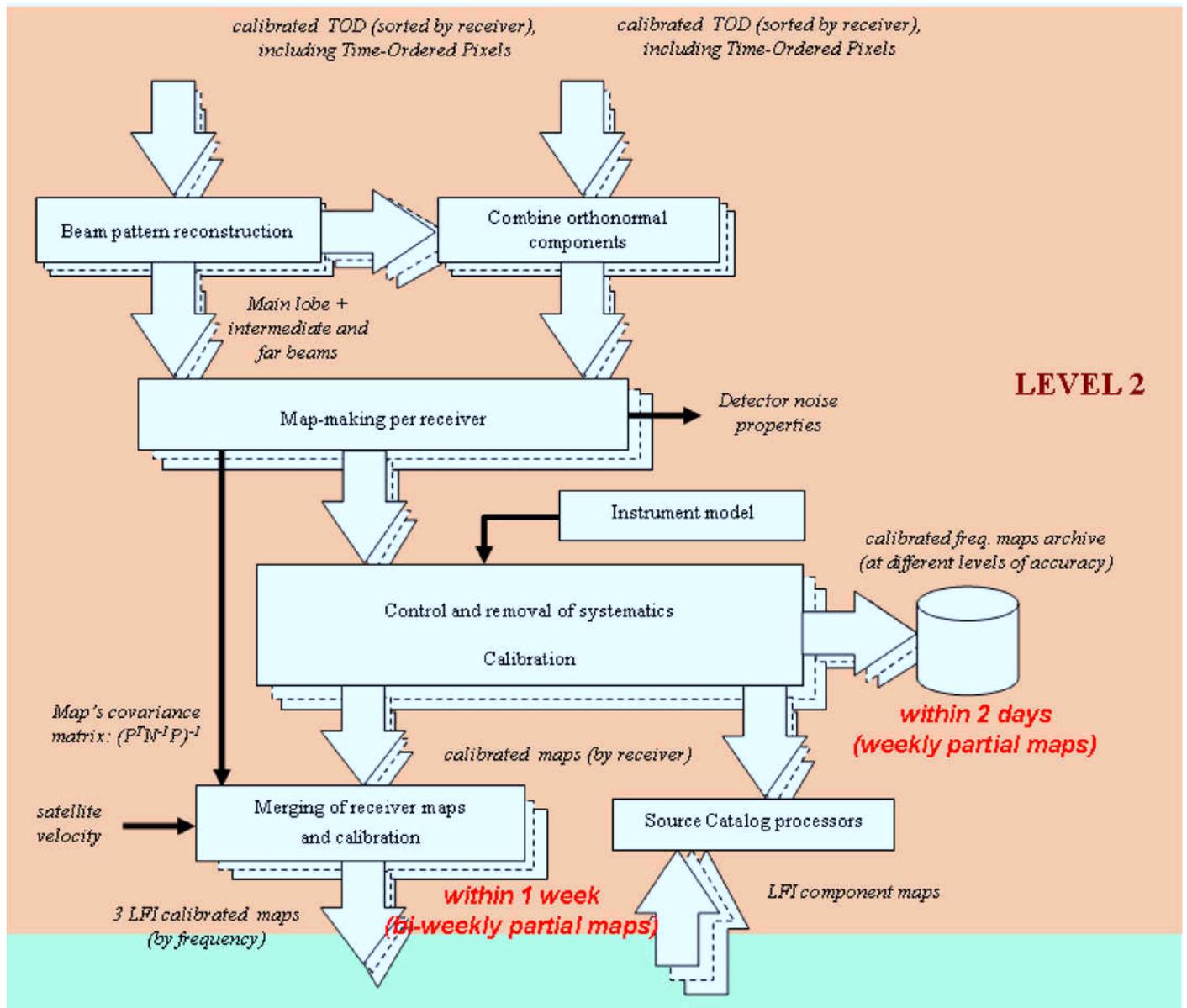}
    \caption{Level~2 Map-making pipeline.}
    \label{MapMaking}
\end{figure*}

The instrument model allows one to check and control systematic
effects and the quality of the removal performed by map-making and
calibration of the receiver map. Receiver maps cleaned of
systematic effects at different levels of accuracy will be stored
into a calibrated map archive. The production of
frequency-calibrated maps will be performed by processing together
all receivers from a given frequency channel in a single
map-making run. In Figs.~\ref{Calibration} and \ref{MapMaking}, we
report the steps performed by Level~2, together with the
associated times foreseen.

\subsection{DPC Level~3}
\label{sec:LFIDPCLevel3}

\begin{figure*}
    \centering
    \includegraphics[width=1.\hsize]{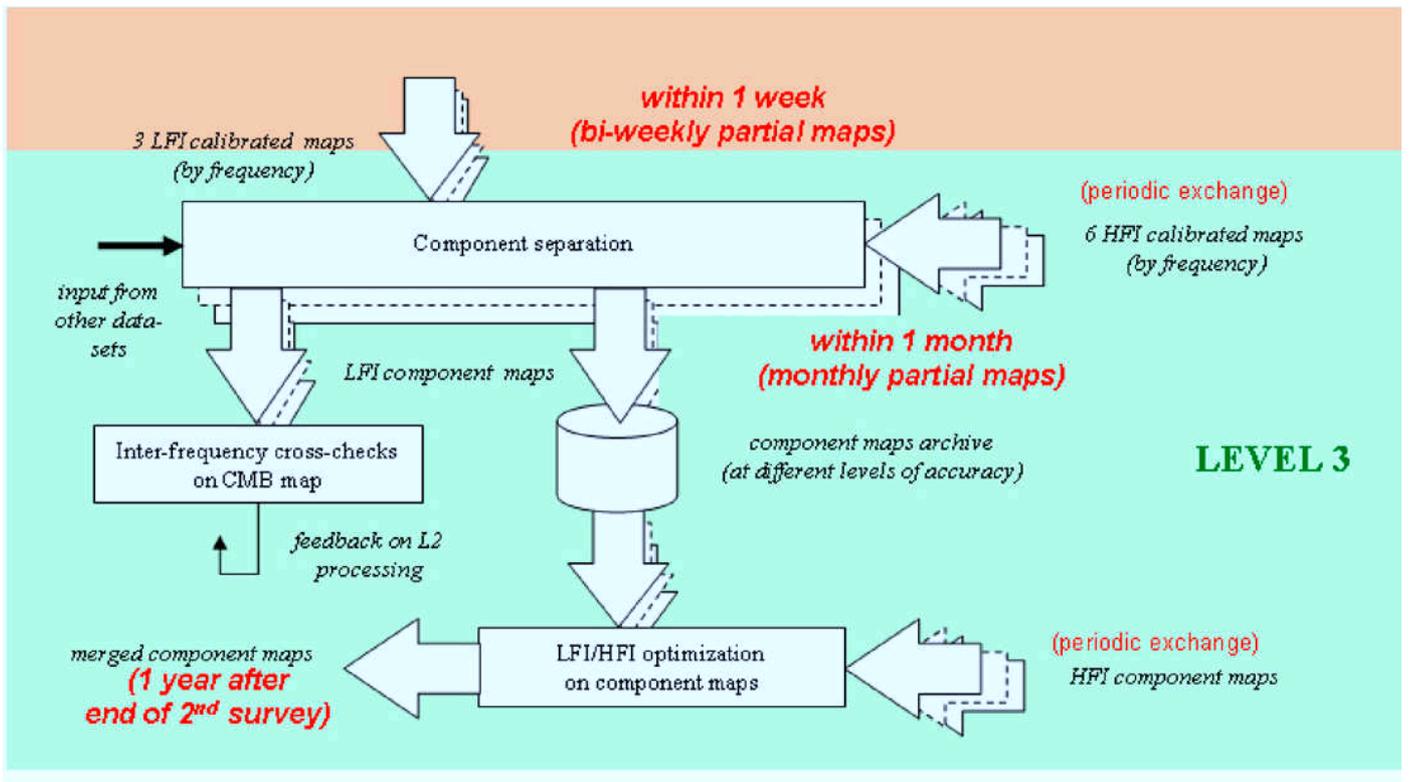}
    \caption{Level~3 pipeline structure.}
    \label{Level3structure}
\end{figure*}
The goal of the DPC Level~3 is to estimate and characterise maps
all the different astrophysical and cosmological sources of
emission (``components") present at {\it Planck\/} wavelengths.
Using the CMB component obtained after point-source extraction and
cleaning from diffuse, Galactic emission, the angular power
spectrum of the CMB is estimated for temperature, polarisation,
and cross temperature/polarisation modes.

The extraction of the signal from Galactic point-like objects, and
other galaxies and clusters is achieved as a first step, either
using pre-existing catalogues based on non-{\it Planck\/} data, or
filtering the multi-frequency maps with optimal filters to detect
and identify beam-like objects (see \citealt{2009MNRAS.394..510H}
and references therein).

The algorithms dedicated to the separation of diffuse emission
fall into four main categories, depending on the criteria
exploited to achieve separation, and making use of the wide
frequency coverage of {\it Planck\/} (see
\citealt{2008A&A...491..597L} and references therein). Internal
linear combination and template fitting achieves linear mixing and
combination of the multi-frequency data with other data sets,
optimized for CMB or foreground recovery. The independent
component analysis works in the statistical domain, without using
foreground modelling or spatial correlations in the data, but
assuming instead statistical independence between the components
that are to be recovered. The correlated component analysis, on
the other hand, makes use of a parametrization of foreground
unknowns, and uses spatial correlations to achieve separation.
Finally, parametric methods consist of modelling foreground and
CMB components by treating each resolution element independently,
achieving fitting of the unknowns and separation by means of a
maximum likelihood analysis. The LFI DPC Level~3 includes
algorithms that belong to each of the four categories outlined
above. The complementarity of different methods for different
purposes, as well as the cross-check on common products, are
required to achieve reliable and complete scientific products.

As for power spectrum estimation, two independent and
complementary approaches have been implemented (see
\citealt{bolpol}, and references therein): a Monte-Carlo method
suitable for high multipoles (based on the {\sc master} approach,
but including cross-power spectra from independent receivers); and
a maximum likelihood method for low multipoles. A combination of
the two methods will be used to produce the final estimation of
the angular power spectrum from LFI data, before its combination
with HFI data. In Fig.~\ref{Level3structure}, we report the steps
performed in the Level~3 pipeline with the associated timescales
foreseen.

The inputs to the Level~3 pipeline are the three calibrated
frequency maps from LFI together with the six calibrated HFI
frequency maps that should be exchanged on a monthly basis. The
Level~3 pipeline has links with most of the stages of the Level~1
and Level~2 pipelines, and therefore the most complete and
detailed knowledge of the instrumental behaviour is important for
achieving its goals. Systematic effects appearing in the
time-ordered data, beam shapes, band width, source catalogues,
noise distribution, and statistics are examples of important
inputs to the Level~3 processing. Level~3 will produce source
catalogues, component maps, and CMB power spectra that will be
delivered to the {\it Planck\/} Legacy Archive (PLA), together
with other information and data needed for the public release of
the {\it Planck\/} products.

\subsection{DPC Level S}
\label{sec:LFIDPCLevelS}

It was widely agreed within both consortia that a software system
capable of simulating the instrument footprint, starting from a
predefined sky, was indispensable for the full period of the {\it
Planck\/} mission. Based on that idea, an additional processing
level, Level S, was developed and upgraded whenever the knowledge
of the instrument improved \citep{Rei06}. Level S now incorporates
all the instrument characteristics as they were understood during
the ground test campaign. Simulated data were used to evaluate the
performance of data-analysis algorithms and software against the
scientific requirements of the mission and to demonstrate the
capability of the DPCs to work using blind simulations that
contain unknown parameter values to be recovered by the data
processing pipeline.

\subsection{DPC software infrastructure}
\label{sec:LFIDPCInfrastructure}

During the entire {\it Planck\/} project, it has been (and will
continue to be) necessary to deal with aspects related to
information management, which pertain to a variety of activities
concerning the whole project, ranging from instrument information
(e.g., technical characteristics, reports, configuration control
documents, drawings, public communications) to software
development/control (including the tracking of each bit produced
by each pipeline). For this purpose, an Integrated Data and
Information System (IDIS) was developed. IDIS \citep{Ben_2000} is
a collection of software infrastructure for supporting the {\it
Planck\/} Data Processing Centres in their management of large
quantities of software, data, and ancillary information. The
infrastructure is relevant to the development, operational, and
post-operational phases of the mission.

\begin{figure*}
    \centering
    \includegraphics[width=1.\hsize]{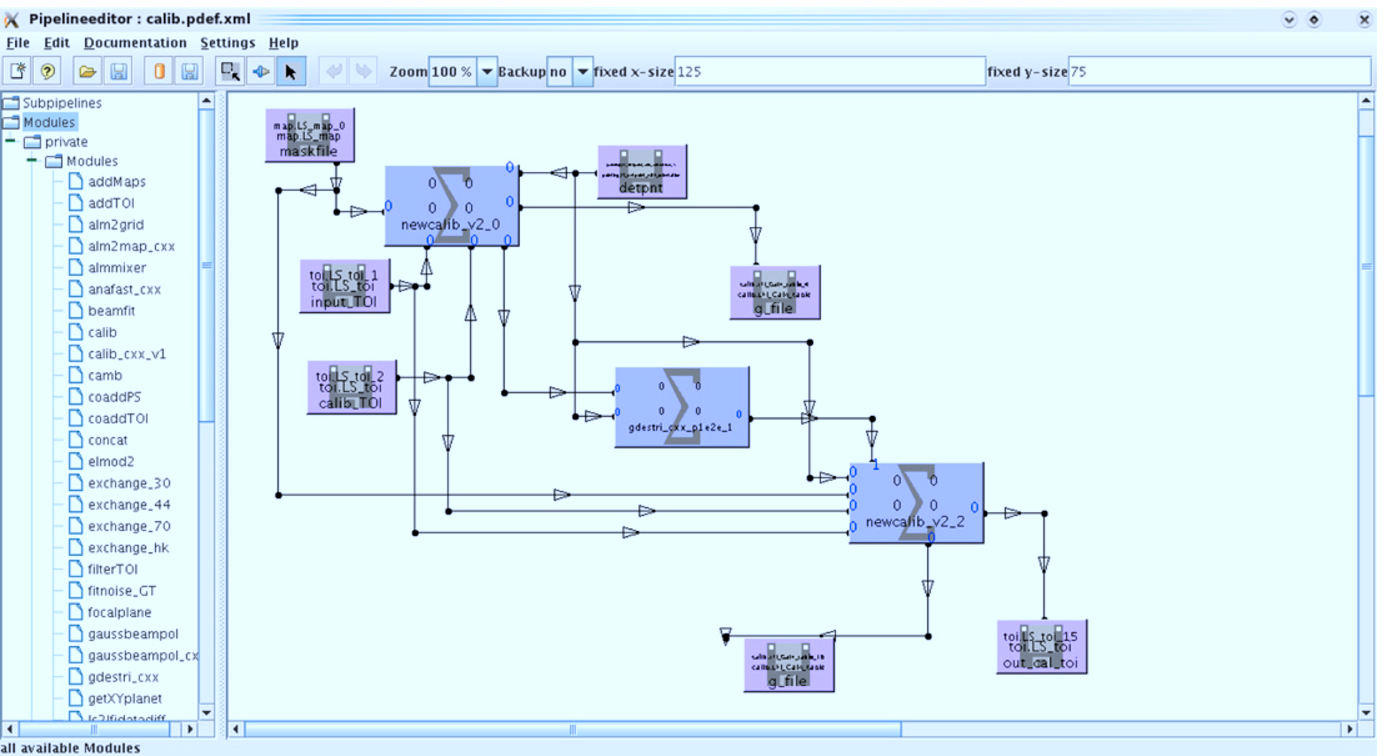}
    \caption{IDIS ProC pipeline editor.}
    \label{IDISproc}
\end{figure*}

The full IDIS can be broken down into five major components:
\begin{itemize}
    \item Document management system (DMS), to store and share documents.
    \item Data management component (DMC), allowing the ingestion, efficient
management, and extraction of the data (or subsets thereof)
produced by {\it Planck\/} activities.
    \item Software component (SWC), allowing the system to administer,
document, handle, and keep under configuration control the
software developed within the {\it Planck\/} project.
    \item Process Coordinator (ProC), allowing the creation and running of
processing pipelines inside a predefined and well controlled environment.
    \item Federation layer (FL), which allows controlled access to the
previous objects and acts as a glue between them.
 \end{itemize}
 The use of the DMS has allowed the entire consortia to ingest and store
hundreds of documents and benefit from an efficient way of
retrieving them. The DMC is an API (application programming
interface) for data input/output, connected to a database (either
relational or object-oriented) and aimed at the archiving and
retrieval of data and the relevant meta-information; it also
features a user GUI. The ProC is a controlled environment in which
software modules can be added to create an entirely functional
pipeline. It stores all the information related to versioning of
the modules used, data, and temporary data created within the
database while using the DMC API. In Fig.~\ref{IDISproc}, an
example of the LFI pipeline is shown. Finally, the FL is an API
that, using a remote LDAP database, assigns the appropriate
permission to the users for data access, software access, and
pipeline run privileges.

\subsection{DPC test performed}
\label{sec:LFIDPCTestPerformed}

Each pipeline and sub-pipeline (Level~1, Level~2, and Level~3) has
undergone different kinds of tests. We report here only the
official tests conducted with ESA, without referring to the
internal tests that were dedicated to DPC subsystems. Level~1 was
the most heavily tested, as this pipeline is considered
launch-critical. As a first step, it was necessary to validate the
output with respect to the input; to do that, we ingested inside
the instrument a well known signal as described in \citet{Fra09}
with the purpose of verifying whether the processing inside
Level~1 was correct. This also had the benefit of providing an
independent test of important functionalities for the REBA
software responsible for the onboard preprocessing of scientific
data. Afterwards, more complete tests, including all interfaces
with other elements of the ground segment, were performed. Those
tests simulate one week of nominal operations (SOVT1 -- system
operation validation test; \citealt{Kec08}) and, during the SOVT2,
one week of the commissioning performance verification (CPV)
phase. During these tests, it was demonstrated that the LFI
Level~1 is able to deal with the telemetry as it would be acquired
during operations.

Tests performed on Level~2 and Level~3 were more science-oriented
to demonstrate the scientific adequacy of the LFI DPC pipeline,
i.e.,~its ability to produce scientific results commensurate with
the objectives of the {\it Planck\/} mission. These tests were
based on blind simulations of growing complexity. The Phase 1 test
data, produced with Level S, featured some simplifying
approximations:
\begin{itemize}
    \item the sky model was based on the ``concordance model'' CMB
    (no non-Gaussianity);
    \item the dipole did not include modulations due to the Lissajous orbit
 around L2;
    \item Galactic emission was obtained assuming non-spatially
    varying spectral index;
    \item the detector model was ``ideal'' and did not vary with time;
    \item the scanning strategy was ``ideal'' (i.e., no gaps in the
data).
 \end{itemize}
The results of this test were in line with the objectives of the mission
\citep[see][]{Per07}).

The Phase 2 tests are still ongoing. They take into account more
realistic simulations with all the known systematics and known
problems (e.g., gaps) in the data.

\section{Pre-launch status}
\label{sec:conc}

We have provided an overview of the Low Frequency Instrument (LFI)
programme and of its organization within the ESA {\it Planck\/} mission.
After a brief description of the {\it Planck\/} main properties and observational strategy, 
the main scientific goals have been presented, ranging from fundamental
cosmology to Galactic and extragalactic astrophysics by focusing on those more relevant to LFI. The LFI  design and development have been
outlined, together with the model philosophy and testing strategy. The LFI
approach to on-ground and in-flight calibration and the LFI ground segment have been described.
We have reported on the data analysis pipeline that has been successfully tested. 

Ground testing shows that the LFI operates as anticipated.  
The observational program will begin
after the {\it Planck/Herschel\/} launch on May 14th, 2009.

A challenging commissioning and final calibration phase will prepare the
LFI for nominal operations that will start about 90 days after launch.
After $\sim 20$ days, the instrument will be switched on and its
functionality will be tested in parallel with the cooling of the $20\,$K
stage. Then the cooling period of the HFI focal plane to $4\,$K will be
used by the LFI to tune voltage biases of the front end amplifiers, phase
switches, and REBA parameters, which will set the final scientific performance
of the instrument.
Final tunings and calibration will be performed in parallel with HFI
activities for about 25 days until the last in-flight calibration phase, the
so-called ``first light survey''.  This will involve 14 days of data acquisition in
nominal mode that will benchmark the whole system, from satellite and
instruments to data transmission, ground segment, and data processing
levels.

The first light survey will produce the very first {\it Planck\/} maps.
This will not be designed for scientific exploitation but will rather serve
as a final test of the instrumental and data processing capabilities
of the mission.
After this, the {\it Planck\/} scientific operations will begin.

{\it Note that at the time of publishing this article, {\it Planck\/}  was launched successfully 
with {\it Herschel}  on May 14th, 2009, and it is in the process of completing its first full sky survey
as foreseen.}  

\section{Acknowledgements}

{\it Planck\/} is a project of the European Space Agency with instruments funded 
by ESA member states, and with special contributions from Denmark and 
NASA (USA). 
The {\it Planck}-LFI project is developed by an International Consortium led
by Italy and involving Canada, Finland, Germany, Norway, 
Spain, Switzerland, UK and USA. 
The Italian contribution to {\it Planck\/} is supported by the Agenzia Spaziale Italiana
(ASI) and INAF.
We also wish to thank the many people of the {\it Herschel/Planck\/} Project
and RSSD of ESA, ASI, THALES Alenia Space Industries and the LFI Consortium
that have contributed to the realization of LFI.
We are grateful to our HFI colleagues for such a fruitful collaboration during
so many years of common work.
The German participation at the Max-Planck-Institut f\"ur Astrophysik is 
funded by the Bundesministerium f\"ur Wirtschaft und Technologie through the 
Raumfahrt-Agentur of the Deutsches Zentrum f\"ur Luft- und Raumfahrt (DLR) 
[FKZ: 50 OP 0901] and by the Max-Planck-Gesellschaft (MPG). 
The Finnish contribution is supported by the Finnish Funding Agency for
Technology and Innovation (Tekes) and the Academy of Finland.
The Spanish participation is funded by Ministerio de Ciencia e Innovacion
through the project ESP2004-07067-C03 and AYA2007-68058-C03.
The UK contribution is supported by the Science and Technology Facilities 
Council (STFC).
C.~Baccigalupi and F.~Perrotta acknowledge partial support of the NASA LTSA
Grant NNG04CG90G.
We acknowledge the use of the BCX cluster at CINECA under the agreement
INAF/CINECA. We acknowledge the use of the Legacy Archive for Microwave
Background Data Analysis (LAMBDA). 
Support for LAMBDA is provided by the NASA Office of Space Science.
We acknowledge use of the HEALPix \citep{2005ApJ...622..759G} software and
analysis package for deriving some of the results in this paper. 

\appendix

\section{List of Acronyms}
\label{acronyms}



\noindent
AIV = Assembly Integration and Verification

\noindent
API = Application Programming Interface

\noindent
APS = Angular Power Spectrum

\noindent
ASI = Agenzia Spaziale Italiana (Italian Space Agency)

\noindent
ATCA = Australian Telescope Compact Array

\noindent
AVM = AVionics Model

\noindent
BEM = Back-End Module

\noindent
BEU = Back-End Unit


\noindent
CDM = Cold Dark Matter

\noindent
{\it COBE\/} = COsmic Background Explorer

\noindent
COBRAS = COsmic Background Radiation Anisotropy Satellite

\noindent
CMB = Cosmic Microwave Background

\noindent
CPV = Commissioning Performance Verification

\noindent
CSL = Centre Spatial de Liege

\noindent
DAE = Data Acquisition Electronics

\noindent
DBI = Dirac-Born-Infeld (inflation)

\noindent
DC = Direct Current

\noindent
DDS = Data Distribution System

\noindent
DMC = Data Management Component

\noindent
DMS = Document Management System

\noindent
DPC = Data Processing Centre

\noindent
EBB = Elegant BreadBoarding

\noindent
EMC = ElectroMagnetic Compatibility

\noindent
ESA = European Space Agency

\noindent
ESOC = European Space Operations Centre

\noindent
ET = Edge Taper

\noindent
FEM = Front-End Module

\noindent
FL = Federation Layer

\noindent
FM = Flight Model

\noindent
FPU = Focal Plane Unit


\noindent
FWHM = Full Width Half Maximum


\noindent
GLAST = Gamma-ray Large  Area Space Telescope

\noindent
GLS = Generalized Least Squares

\noindent
GSE = Ground Support Equipment

\noindent
GUI =  Graphical User Interface

\noindent
HEALPix = Hierarchical Equal Area isoLatitude Pixelization

\noindent
HEMT = High Electron Mobility Transistor

\noindent
HFI = High Frequency Instrument

\noindent
HPST = High-Pressure Stabilization Tank


\noindent
IDIS = Integrated Data and Information System


\noindent
IR = Infra Red

\noindent
ISM = Inter-Stellar Medium

\noindent
JPL = Jet Propulsion Laboratory

\noindent
JT = Joule-Thomson

\noindent
LDAP = Lightweight Directory Access Protocol

\noindent
LFI = Low Frequency Instrument

\noindent
LIFE = LFI Integrated perFormance Evaluator

\noindent
LNA = Low Noise Amplifier

\noindent
LPSB = Low-Pressure Storage Bed

\noindent
LVHX = Liquid Vapour Heat eXchange

\noindent
MIB = Mission Information Base

\noindent
MIC = Microwave Integrated Circuit

\noindent
MMIC = Monolithic Microwave Integrated Circuit

\noindent
MOC = Mission Operation Centre

\noindent
NASA = National Aeronautics and Space Administration (USA)

\noindent
NG = Non Gaussianity

\noindent
OMT = Orthomode Transducer

\noindent
PACE = Piping and Cold-End Assembly

\noindent
PD = Prototype Demonstrator

\noindent
PFM = {\it Planck\/} ProtoFlight Model


\noindent
PI = Principal Investigator

\noindent
PID = Proportional Integral Derivative


\noindent
PLA = {\it Planck\/} Legacy Archive

\noindent
PPLM = {\it Planck\/} PayLoad Module

\noindent
ProC = Process Coordinator

\noindent
PS = Phase Switch

\noindent
QM = Qualification Model


\noindent
RAA = Radiometer Array Assembly

\noindent
RCA = Radiometer Chain Assembly

\noindent
REBA = Radiometer Electronics Box Assembly

\noindent
RF = Radio Frequency

\noindent
RTA = Real Time Assessment

\noindent
SAMBA = SAtellite for Measurement of Background Anisotropies

\noindent
SCE = Sorption Cooler Electronics

\noindent
SCS = Sorption Cooler Subsystem

\noindent
SOVT = System Operation Validation Test

\noindent
SS = Scanning Strategy

\noindent
SVM = SerVice Module

\noindent
SWC = SoftWare Component

\noindent
TM = TeleMetry

\noindent
TMU = Thermo-Mechanical Unit

\noindent
TOI = Time Order Information

\noindent
TQL = Telemetry Quick Look

\noindent
TSA = Temperature Stabilization Assembly; Time Series Analysis

\noindent
{\it WMAP\/} = Wilkinson Microwave Anisotropy Probe

\noindent
WR = Warm Radiator

\noindent
XPD = Cross-Polar Discrimination

\bibliography{LFI_Programme_Paper_30jun09_revised_referee_editorial}  

\end{document}